\documentclass[iop]{emulateapj}

\usepackage{apjfonts}

\newcommand{\Sersic}{S\'{e}rsic }
\def\lae{\mathrel{\raise .4ex\hbox{\rlap{$<$}\lower 1.2ex\hbox{$\sim$}}}}
\def\gae{\mathrel{\raise .4ex\hbox{\rlap{$>$}\lower 1.2ex\hbox{$\sim$}}}}

\shorttitle{Pictor A Jet and Tidal Tail}
\shortauthors{Gentry et al.}

\begin{document}
\title{Optical Detection of the Pictor A Jet and Tidal Tail: Evidence against an IC/CMB jet}
\author{Eric S.\ Gentry\altaffilmark{1,2},
Herman L. Marshall\altaffilmark{2},
Martin J. Hardcastle\altaffilmark{3},
Eric S. Perlman\altaffilmark{4},
Mark Birkinshaw\altaffilmark{5,6}, 
Diana M. Worrall\altaffilmark{5,6}, 
Emil Lenc\altaffilmark{7,8}, 
Aneta Siemiginowska\altaffilmark{6}, 
C. Megan Urry\altaffilmark{9}
}
\altaffiltext{1}{Dept. of Astronomy and Astrophysics,
University of California at Santa Cruz, 1156 High St., Santa Cruz, CA, 95064, USA; egentry@ucsc.edu}
\altaffiltext{2}{Kavli Institute for Astrophysics and Space Research,
Massachusetts Institute of Technology, 77 Massachusetts Ave.,
Cambridge, MA 02139, USA}
\altaffiltext{3}{School of Physics, Astronomy, and Mathematics,
University of Hertfordshire, College Lane, Hatfield, Hertfordshire
UK AL10 9AB}
\altaffiltext{4}{Dept. of Physics and Space Sciences,
Florida Institute of Technology, 150 W. University Blvd., Melbourne, FL, 32901, USA}
\altaffiltext{5}{Dept. of Physics, University of Bristol, Tyndall Ave., Bristol BS8 1TL, UK}
\altaffiltext{6}{Harvard-Smithsonian Center for Astrophysics,
60 Garden St., Cambridge, MA 02138, USA}
\altaffiltext{7}{Sydney Institute for Astronomy, School of Physics, The University of Sydney, NSW 2006, Australia}
\altaffiltext{8}{ARC Centre of Excellence for All-sky Astrophysics (CAASTRO)}
\altaffiltext{9}{Department of Physics, Yale University, New Haven, CT 06511, USA }
\slugcomment{Received 2015 May 9; Accepted 2015 June 12}

\begin{abstract}

New images from the Hubble Space Telescope of the FRII radio galaxy
Pictor A reveal a previously undiscovered tidal tail, as well as a number
of jet knots coinciding with a known X-ray and radio jet.
The tidal tail is approximately 5\arcsec\ wide (3 kpc projected), starting
18\arcsec\ (12 kpc) from the center of Pictor A, and extends more than
90\arcsec\ (60 kpc).  The knots are part of a jet observed to be about
4\arcmin\ (160 kpc) long, extending to a bright hotspot.
These images are the first optical detections of this jet, and by extracting
knot flux densities through three filters we set constraints on emission models.
While the radio and optical flux densities are usually explained by 
synchrotron emission, there are several emission mechanisms
which might be used to explain the X-ray flux densities.
Our data rule out Doppler boosted inverse Compton scattering
as a source of the high energy emission.
Instead, we find that the observed emission can be well described by
synchrotron emission from electrons with a 
low energy index ($p\sim2$) that dominates the radio band,
while a high energy index ($p\sim3$) is needed for the X-ray band
and the transition occurs in the optical/infrared band.
This model is consistent with a continuous electron injection scenario.
\end{abstract}


\section{Introduction}

To date more than 40
extragalactic jets have detections in the IR-UV
band although the
majority of these are nearby, low-power FR I jets, where the speeds are
slower, and processes and environments are different from those of
quasars and FRII radio galaxies.  
While many FR II quasar jets have been detected in HST images, most are so
distant that only a few point-like emission regions are found.
Resolving structure in the knots is rare; one
example of resolved structure is 3C 273, which has an unusually bright
jet in the optical band and is close enough that HST resolves
each knot \citep{2001ApJ...549L.167M,2006ApJ...648..900J}.
The proximity of Pictor A and its FR II jet (located at a redshift of $z=0.035$)
makes it an appealing candidate for resolving small, dim features
not detectable in more distant quasars.
Observing these features would improve our understanding of the physical processes
active within these jets.

There is currently some uncertainty regarding the primary emission mechanisms
active within these jets at kiloparsec scales. Observations by the {\em Hubble} Space Telescope (HST) 
have shown that often the radio and X-ray spectra cannot be connected smoothly via a
single synchrotron model.  
This could be explained by a jet with relativistic bulk motion,
where the high-energy emission originates from relativistically \emph{boosted} inverse Compton scattering
of cosmic microwave background photons
\citep[IC/CMB; proposed by][]{2000ApJ...544L..23T,2001MNRAS.321L...1C}.
Boosted IC/CMB emission has the attractive consequence of allowing
radio observations to help constrain the conditions and environment of the
X-ray emitting particles; no additional electron populations need to be invoked.
When lacking better data, surveys have also assumed boosted IC/CMB emission
to constrain the jet geometry 
\citep[who tentatively classified Pictor A as a boosted jet]{Marshall2005,2002ApJ...565..244H}.
The IC/CMB model also was supported by data from a number of early surveys.
A majority of the systems studied by \citet{2004ApJ...608..698S} were best explained
by an IC/CMB model, and follow up observations of PKS 1136-135 and 1150+497
by \citet{2006ApJ...641..717S} verified predictions of the IC/CMB model.

Despite this early evidence, the data for a number of these jets 
are no longer consistent with a boosted IC/CMB model.
\citet{2015ApJ...805..154M} used gamma ray observations to 
disprove the boosted IC/CMB model for PKS 0637-572,
which had previously been the prototypical example of
a boosted IC/CMB jet \citep{2000ApJ...544L..23T,2001MNRAS.321L...1C}.
\citet{2041-8205-780-2-L27} also used gamma ray observations
to rule out a relativistically boosted jet in
3C 273, building on the work of \citet{2007MNRAS.380..828J} and \citet{2006ApJ...648..910U}.
The high energy emission from PKS 1136-135, 
one of the jets used by \citet{2006ApJ...641..717S} to verify IC/CMB predictions,
has since been found to have a high level of polarization which is incompatible with an IC/CMB model
\citep{2013ApJ...773..186C}.
The data of all of these systems were previously consistent with
a boosted IC/CMB model (most notably PKS 0637-572),
but better data have ruled out this explanation.
Here we contribute evidence against such a model for Pictor A,
which was also initially labeled a boosted IC/CMB jet.

Pictor A is an FR II radio galaxy at a redshift of $0.035$.
(For $H_0 = 70.5$ km/s/Mpc, at this redshift, 1\arcsec\ corresponds to about 700 pc.)
Pictor A's pencil-like X-ray jet was discovered in {\em Chandra} observations
by \citet{Wilson01}, extending 1\farcm9 from the core,
oriented toward the resolved NW hotspot 4\farcm2 from the core.
The jet is barely visible in radio images due to its low brightness relative to
the core and the extended lobes \citep{Perley97,1999ApJS..123..447S,Marshall10}.
Simkin et al.\ discovered H$\alpha$ emission about 5\arcsec\ from the
core along the direction of the jet to the NW hotspot but found no other
optical emission from the jet.
\cite{Hardcastle05} first reported a weakly-detected 1 keV X-ray counter-jet
extending to the E hotspot; more recent X-ray data of M. J. Hardcastle et al.\ (2015, in preparation)
confirm this detection.

\citet{Wilson01}, on the basis of a boosted IC/CMB model for the
X-rays, showed that the magnetic field in the jet would need to be
substantially below equipartition, $\sim 2 \times 10^{-6}$ G,
assuming that the jet is Doppler boosted with a bulk Lorenz factor $\Gamma = 2$ -- 6 at
$\theta<30$\arcdeg\ to the line of sight.
However, \citet{Hardcastle05} argued that the well-constrained steep X-ray photon
index favors a synchrotron model for the X-rays, in which case the
magnetic field strength and boosting parameters are not well
constrained.

Examining {\em Chandra} X-ray images of the jet, we found
evidence for flares in the jet at $3-4.5\sigma$ significance
\citep{Marshall10}.
The projected size of the jet is over 150 kpc long (4\arcmin) and about a kpc wide, so finding localized
brightness changes on 1-year time scales was surprising.
Using the 5 GHz radio flux density, the equipartition magnetic field is $B_{\rm eq} = 17 \mu$G 
assuming no boosting.
If the 1 keV X-rays are due to synchrotron emission,
then the corresponding electrons have $\gamma  \approx 7 \times 10^7$ (for $E = \gamma m_e c^2$)
and the synchrotron loss time is $\approx$ 1200 yr, similar to the
light-crossing time scale for a 1\arcsec\ sized source: 2000 yr, which is much longer
than the observed variability time scale.
One possible explanation for the variability is
that the emission arises from a very small knot inside the jet, perhaps as small
as 0\farcs002, so HST observations
were needed to investigate whether there are features $\lae$ 0\farcs1 in size.
In addition to investigating these variable regions,
HST data offered the opportunity to isolate and compare SEDs of multiple components within the jet,
which had never been done for Pictor A.

Here, we report the first optical detection of Pictor A's jet and its previously unknown
tidal tail using the HST Wide Field Camera 3 (\emph{WFC3}).
These data were combined with data from the Chandra X-ray Observatory,
as well as the Australia Telescope Compact Array,
in order to obtain a multi-wavelength view of this jet.

\section{Observations and Data Reduction}
\label{section:observations}

Chandra X-ray data (1 keV) were reduced and provided by M. J. Hardcastle et al.\ (2015, in preparation), with
radio (5 GHz) data from \cite{Marshall10}, 
with the process further described in M. J. Hardcastle et al.\ (2015, in preparation).  
In our published images we also
make use of X-ray (1 keV) contours produced from the data used by \cite{Marshall10}, 
which are a subset of the observations by M. J. Hardcastle et al.\ (2015, in preparation).
The rest of this section will focus on the optical data reduction.

Three images with the WFC3
were obtained under the HST Guest Observer program (\emph{GO} proposal ID 12261).
The wideband filters used were:
F160W, F814W, and F475W (see Table~\ref{tab:observations}). 
The F160W image was taken with the infrared detector of WFC3 (\emph{WFC3/IR}), while
the F814W and F475W images were taken with the UV-visible detector (\emph{WFC3/UVIS}).
The raw images were reprocessed using {\tt DrizzlePac} \citep[see][]{2012drzp.book.....G}, providing
images at 0\farcs02 binning for UVIS images and 0\farcs07 binning for the IR image.
In the IR image, the host galaxy is quite bright and extensive, adding significant background
to a large fraction of the image, which requires that it be removed carefully before searching for jet features.

\begin{deluxetable}{lrrrrr}
\tablecolumns{4}
\tablewidth{0pc}
\tablecaption{{\em Hubble}  Observations of the Pictor A Jet \label{tab:observations} }
\tablehead{
     \colhead{Start Date} & \colhead{Filter} & \colhead{$\lambda$\tablenotemark{a}} & \colhead{$\Delta \lambda$\tablenotemark{a}} & \colhead{Exposure} \\
     \colhead{} & \colhead{} & \colhead{(nm)} & \colhead{(nm)} & \colhead{(s)}  }
\startdata
2011 Feb 25 & F160W & 1536 & 268 & 2708  \\
2011 Feb 25 & F814W &  802 & 154 & 1200   \\
2011 Feb 25 & F475W &  477 & 134 & 1299    \\
\enddata
\tablenotetext{a}{Pivot wavelength, as defined in the WFC3 handbook \citep{2015wfci.book.....D}.}
\tablenotetext{a}{Passband rectangular width, as defined in the WFC3 handbook \citep{2015wfci.book.....D}.}
\end{deluxetable}

\subsection{Galaxy Fitting}

\begin{figure*}
\includegraphics[width=15cm]{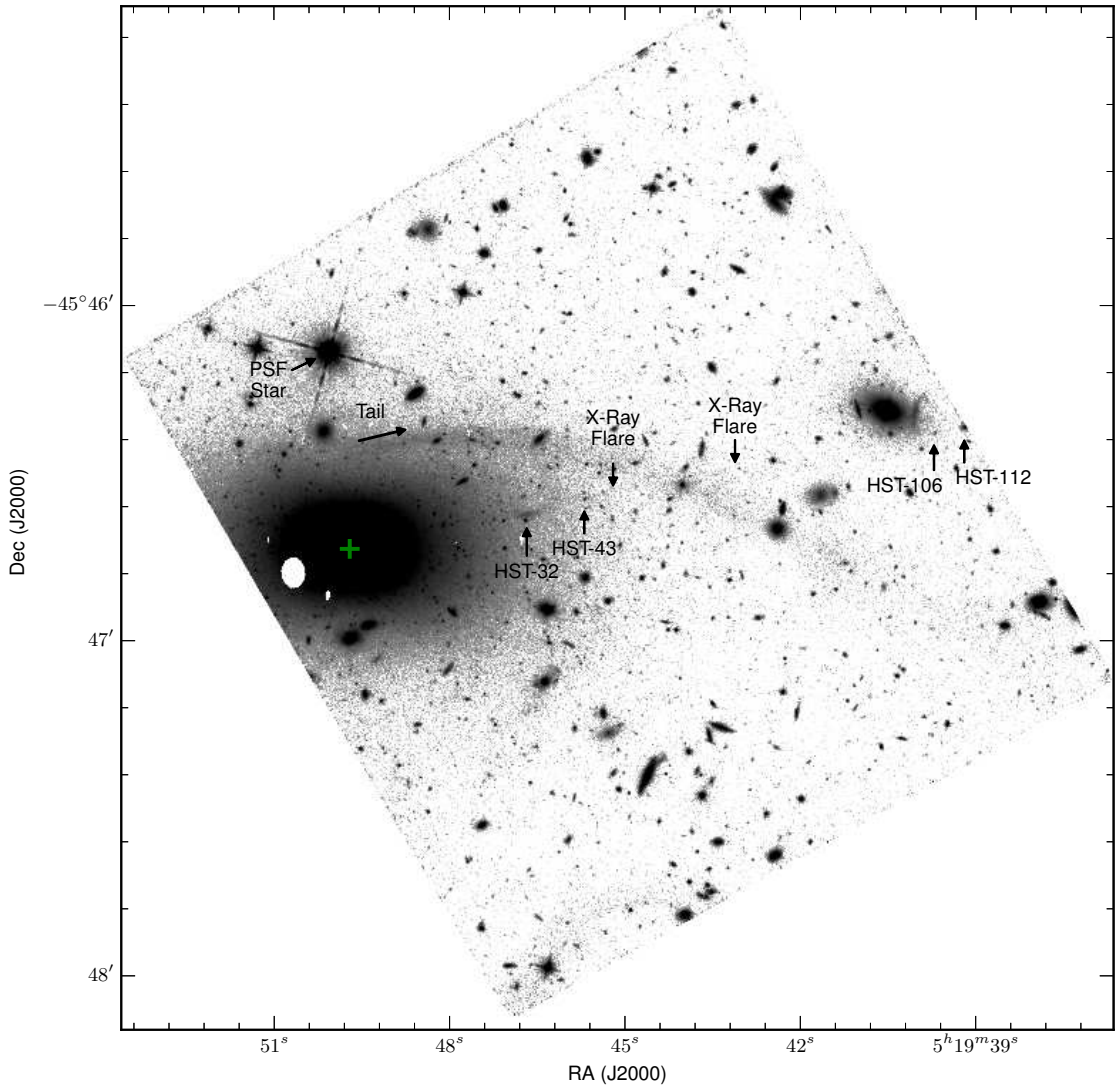}
\caption{
 Image from the WFC3/IR, using filter F160W.  Data were binned to a scale 
 of 0\farcs07, and smoothed with a Gaussian with a FWHM of 0\farcs1. 
 A grayscale was applied with a logarithmic stretch function.
 Central core is marked with a green cross.
 Jet knots HST-32 through HST-112 are labeled, along with the
 locations of the X-ray flares reported by \citet{Marshall10}
 and the star used for a PSF in the galaxy-fitting process.
 The tidal tail is marked by a tangent arrow from the innermost point at which it is detected.
 Bad pixel regions have been masked in white, resulting in the ovals within Pictor A.
 }
\label{fig:f160w:galaxy}
\end{figure*}

The program {\tt galfit} \citep{GALFIT} was used to generate a model of the Pictor A host galaxy and also
generate uncertainty maps. The generated models consisted of a PSF core (caused by the central AGN),
along with 2 generalized \Sersic profiles to account for the elliptical host galaxy.  
The PSF was extracted from a nearby bright star, 
visible to the north of Pictor A and marked in Figure~\ref{fig:f160w:galaxy}.
The \Sersic bulges are described by \Sersic indices $n$ and effective radii $r_e$.
We found that a positive $C_0$ parameter (``boxiness'') was necessary to describe the outer bulge,
but the inner bulge could be sufficiently fit with $C_0=0$.

\begin{figure*}
 \includegraphics[width=15cm]{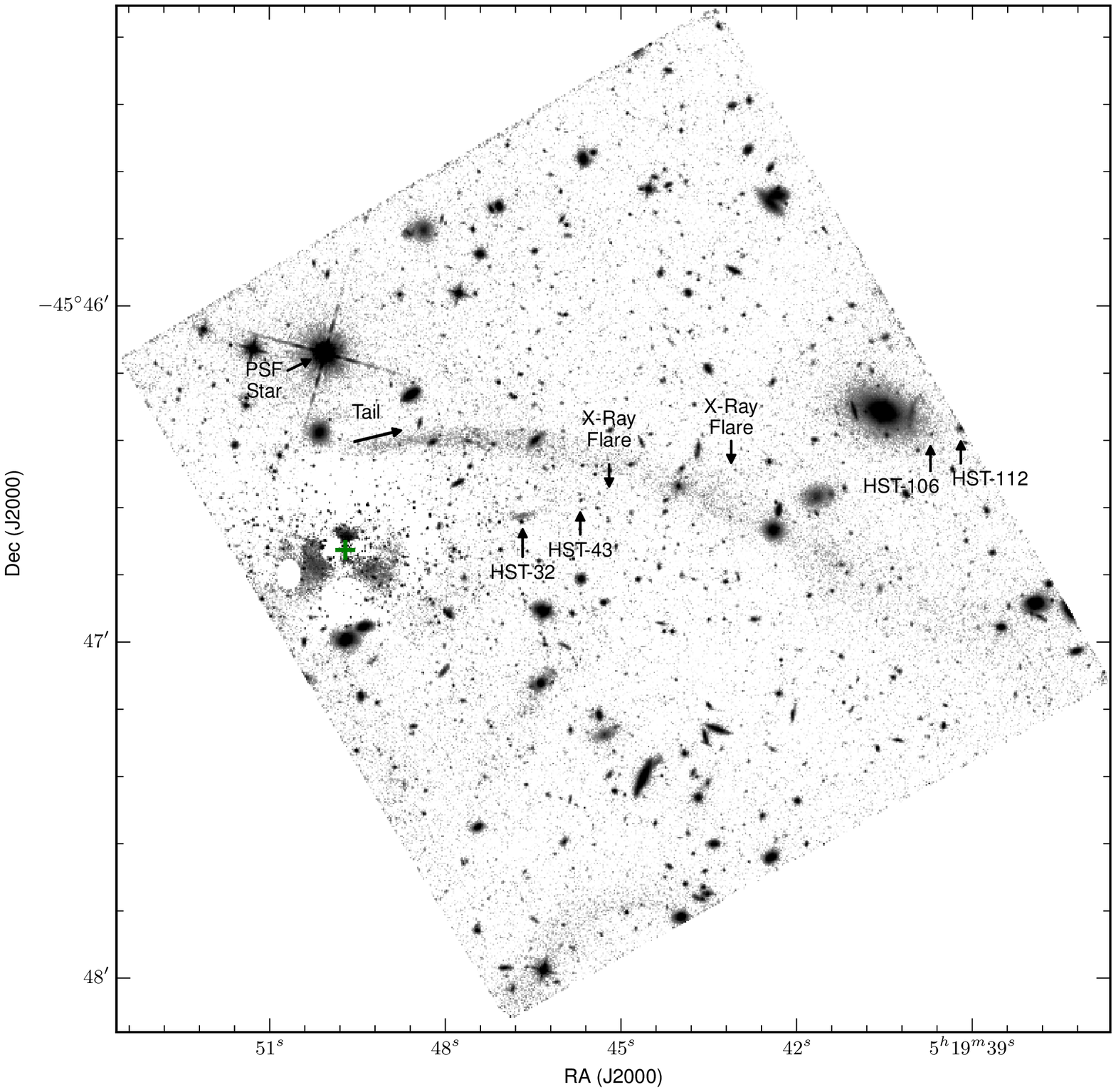}
\caption{
  Same as Figure~\ref{fig:f160w:galaxy}, but now a galaxy-subtracted image 
  from  WFC3/IR, using filter F160W.
  After the subtraction, the jet knots and the innermost extent of the tidal tail are more readily visible.
  The subtraction resulted in strong systematic errors within 15\arcsec\ of the active nucleus.
  }
\label{fig:f160w:resid}
\end{figure*}

In reality we expect the profile to be more complicated,
as the tidal tail provides evidence of a recent merger.  These galaxy fits were primarily used 
to remove the large-scale structure of the host galaxy,
allow us to visually identify knots.
We do not claim any knot detections within 15\arcsec\ of the core
due the rapidly varying fit residuals and the crowding of sources associated with the host galaxy.
In general, we sought the simplest fit
which would still allow for knot identification
and accurate flux extraction.
The galaxy fits do not provide precise photometry for Pictor A.
Results can be seen in Table~\ref{tab:morphologies} and Figures~\ref{fig:f160w:galaxy}-\ref{fig:f160w:resid}. The F814W and F475W images did not require galaxy
subtraction for the knot regions to be clearly separated from the bulge light (see Figure~\ref{fig:jet}).  Figure~\ref{fig:jet} features the jet,
our primary region of interest.

\begin{deluxetable*}{llcccccccc}
\tablecolumns{10}
\tablewidth{0pc}
\tablecaption{Galaxy Fit Parameters \label{tab:morphologies} }
\tablehead{
     \colhead{Filter} & \colhead{\Sersic Component} & \colhead{$m_{\rm core}$\tablenotemark{a}}
         & \colhead{$B_{\rm sky}$\tablenotemark{b}} & \colhead{$m_{\rm bulge}$\tablenotemark{c}} & \colhead{$n$\tablenotemark{d}}
      & \colhead{$r_e$\tablenotemark{e}} & \colhead{$b/a$\tablenotemark{f}}
  & \colhead{$\theta_{\rm PA}$\tablenotemark{g}} & \colhead{$C_0$\tablenotemark{h}} \\
     \colhead{} & \colhead{} & \colhead{} & \colhead{(ADU pix$^{-1}$)}
          & \colhead{}  & \colhead{} & \colhead{(\arcsec)}& \colhead{}
          & \colhead{(\arcdeg)} & \colhead{} }
\startdata
F160W & outer bulge & 17.5 & -3.0 & 16.9 & 2.3 & 16.6 & 0.6 & -91.3 & 0.5 
\\
  &   inner bulge & \nodata & \nodata & 17.4 & 1.6 & 4.2 & 0.8 & -94.6 & 0 
\enddata
\tablenotetext{a}{Apparent (ST) magnitude of the PSF model of the core.}
\tablenotetext{b}{Sky background, assumed to be uniform across the image.}
\tablenotetext{c}{Apparent magnitude of the bulge component.} 
\tablenotetext{d}{\Sersic index.}
\tablenotetext{e}{Effective (half-light) radius.}
\tablenotetext{f}{Ratio of the elliptical \Sersic profile's semi-minor and semi-major axes.}
\tablenotetext{g}{Position angle of the elliptical \Sersic profiles' semi-major axis,
defined to be zero if oriented north-south and increasing as the ellipse is rotated to the east.}
\tablenotetext{h}{The ``diskyness/boxiness'' ratio, which is greater than zero for more box-like isophotes.}
\end{deluxetable*}


\subsection{Knot identification}

\begin{figure*}[tbp]
\centering
\includegraphics[width=\columnwidth]{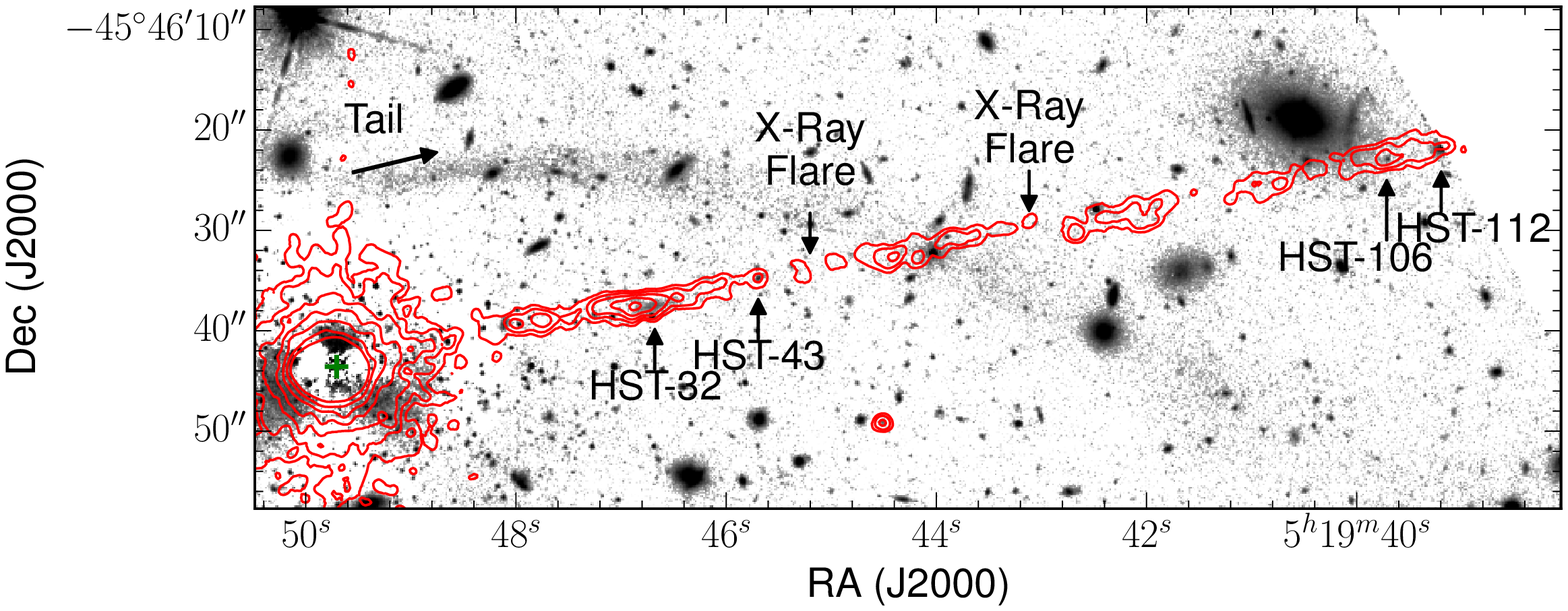}
\includegraphics[width=\columnwidth]{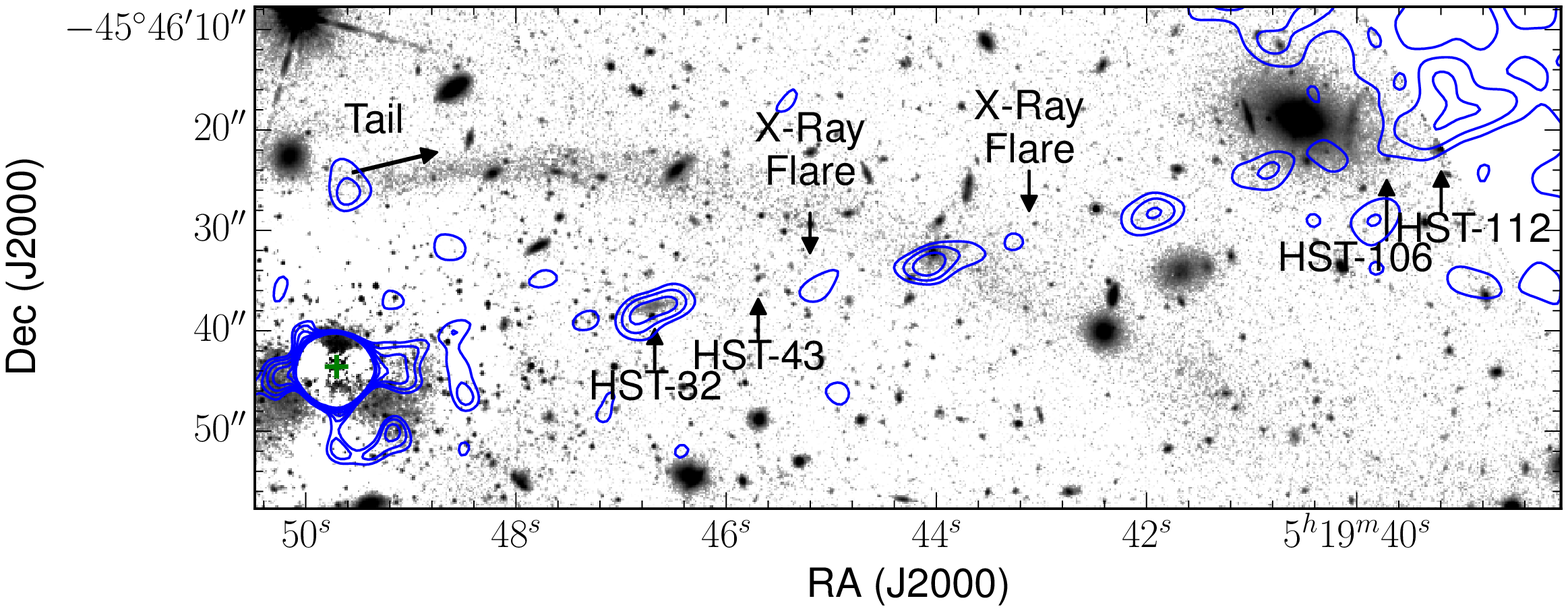}

\includegraphics[width=\columnwidth]{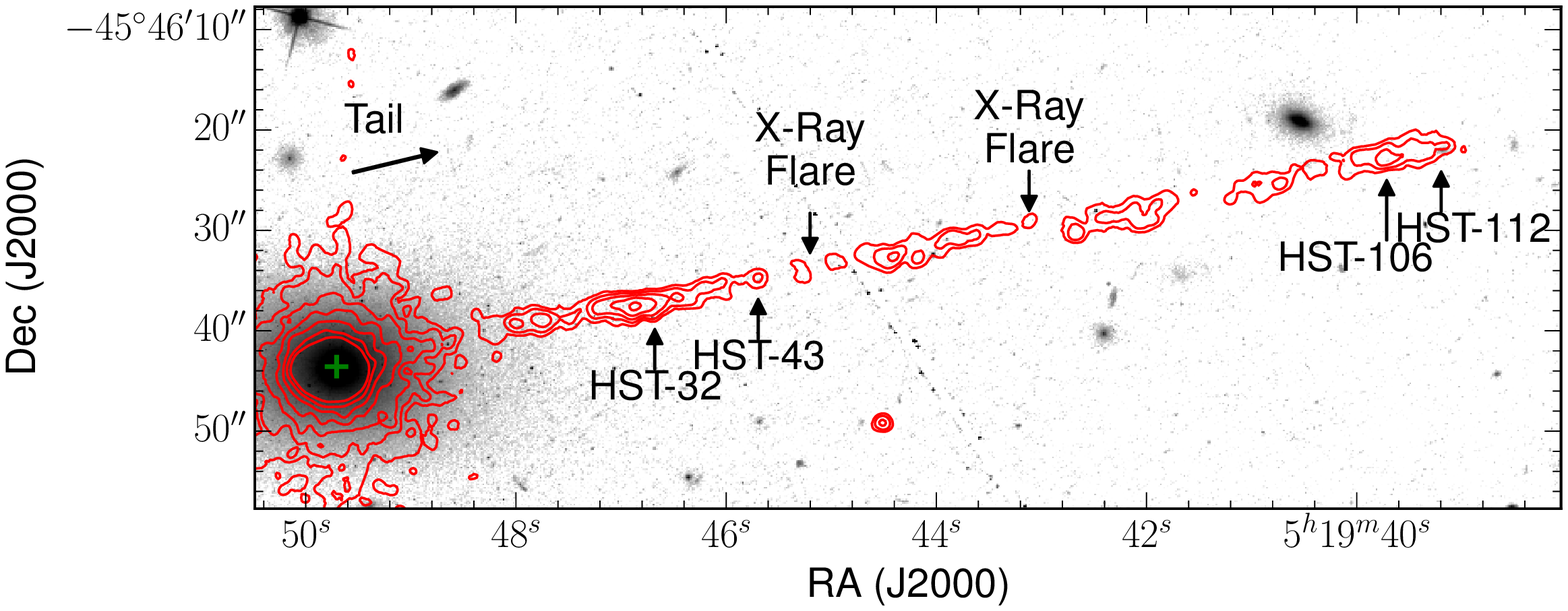}
\includegraphics[width=\columnwidth]{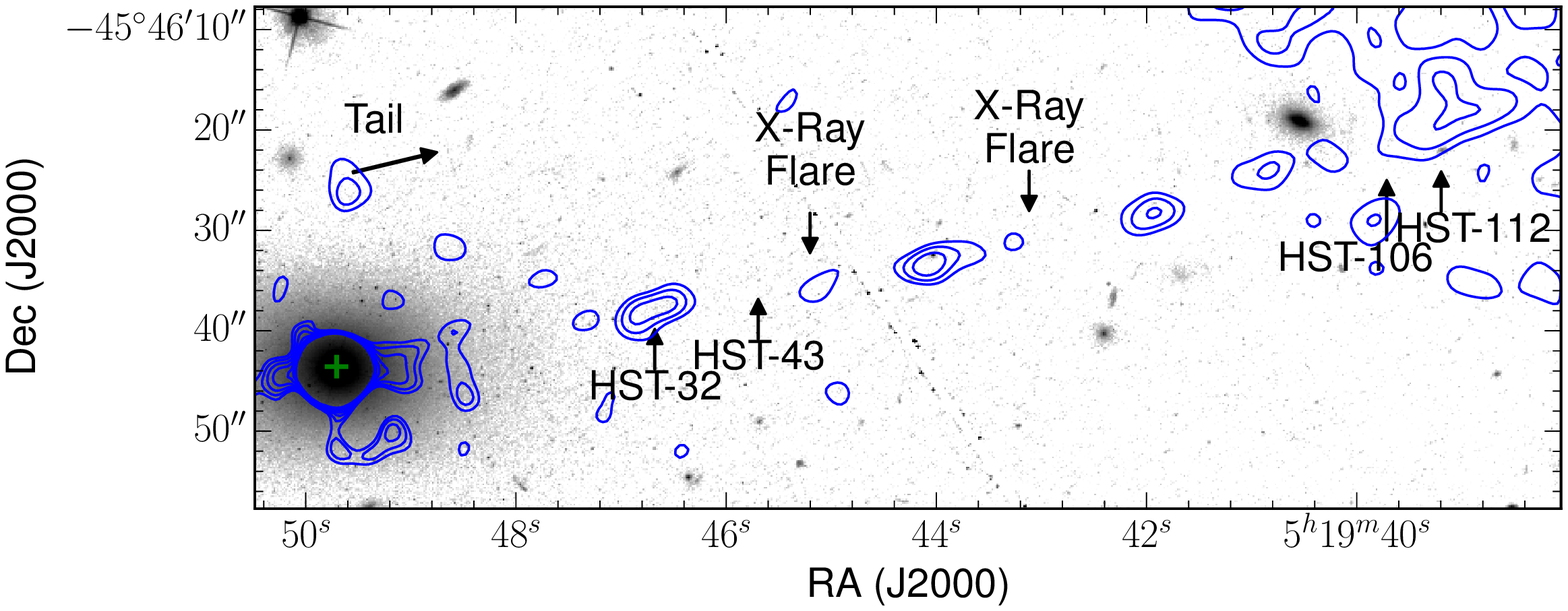}

\includegraphics[width=\columnwidth]{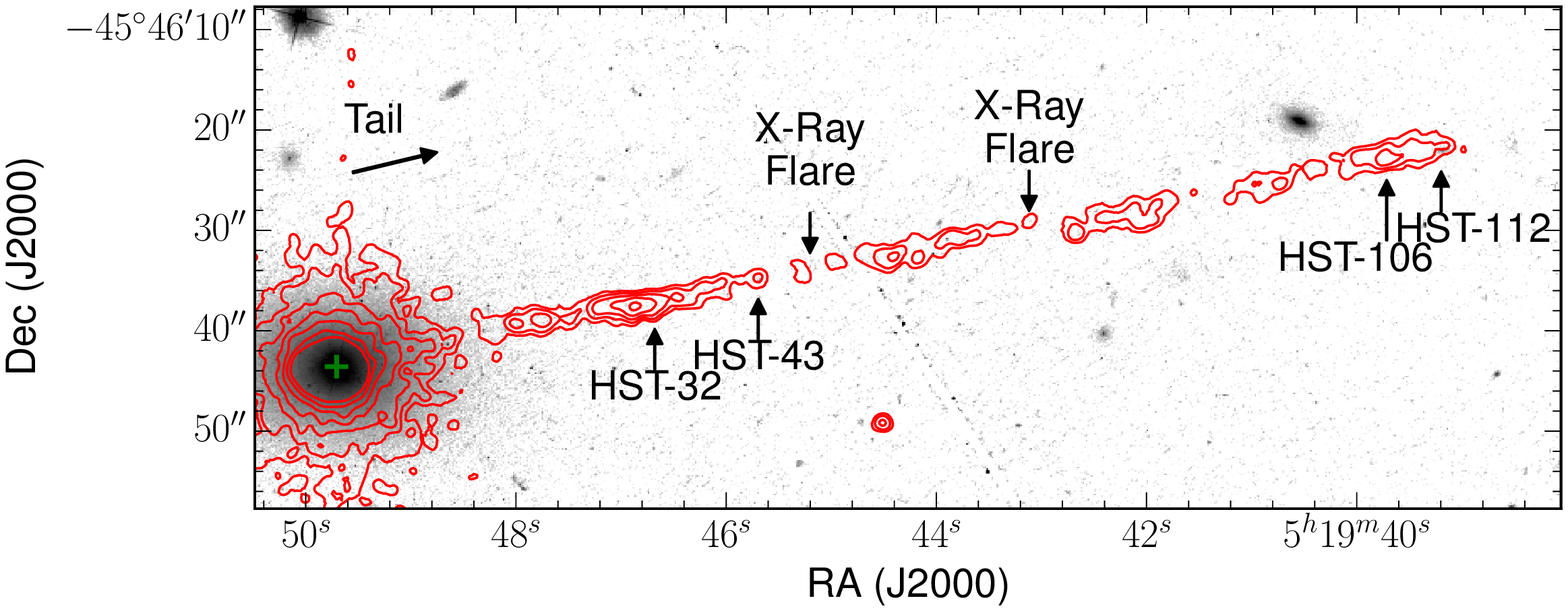}
\includegraphics[width=\columnwidth]{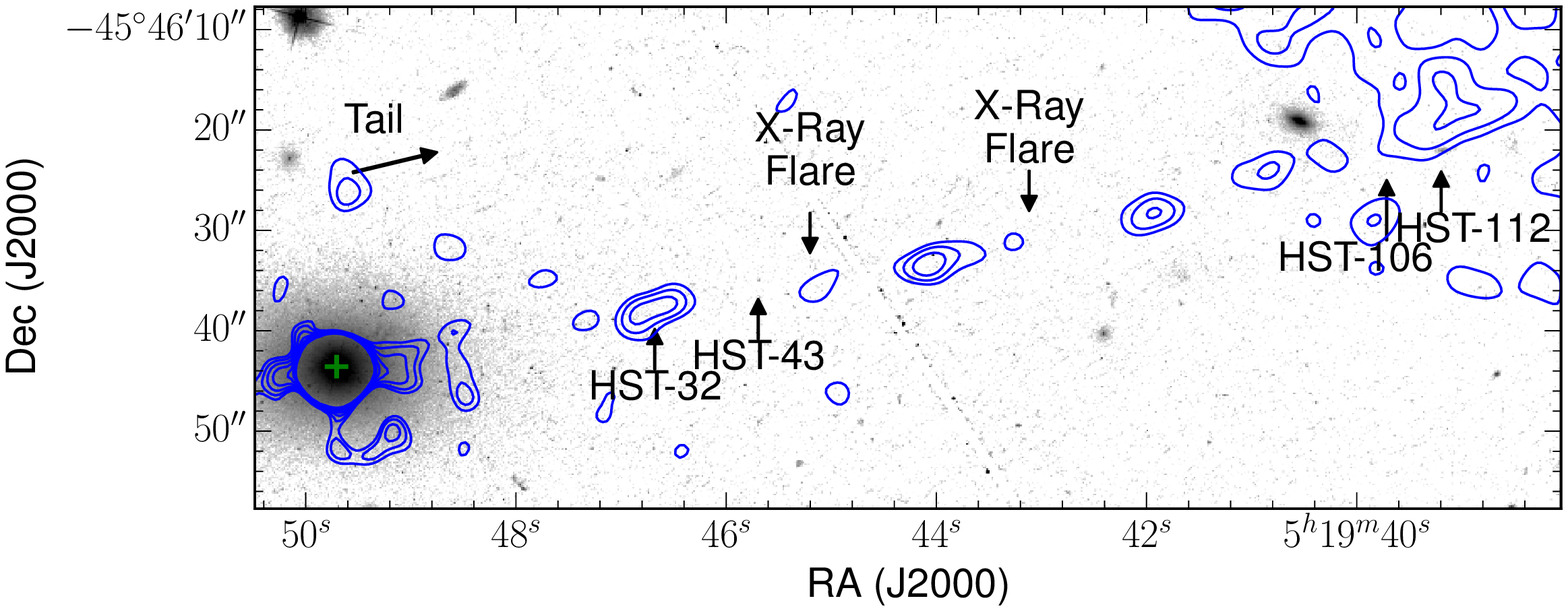}

  \caption{
  The overall F160W image (top) from Figure~\ref{fig:f160w:resid}, 
  cropped to feature the jet,
  along with identical regions from the F814W (middle) and F475W (bottom) images.
  The F160W image has been galaxy-subtracted,
  while the F814W and F475W images have not.
  All images have been smoothed identically, and have similar logarithmic
  stretch functions.
  X-ray contours (left, red) were produced using 7 levels, logarithmically
  spaced between $0.5 - 12.5$  nJy arcsec$^{-2}$.
  Radio contours (right, blue) were also produced using 7 logarithmically
  spaced levels, ranging from $1.25 - 5$ mJy beam$^{-1}$.
  (These contour levels begin approximately at the noise level of each observation.)
  Knots are featured in Figures~\ref{fig:inner_knots}-\ref{fig:outer_knots}.
  }
  \label{fig:jet}
\end{figure*}

Using our images (galaxy-subtracted in the case of the F160W image),
jet knots were identified visually by looking for features coincident
among radio, optical and X-ray bands. 
Figure~\ref{fig:jet} shows the jet's extent in all bands
with X-ray and radio contours overlaid.

\begin{figure*}[tbp]
\centering
\includegraphics[width=\columnwidth]{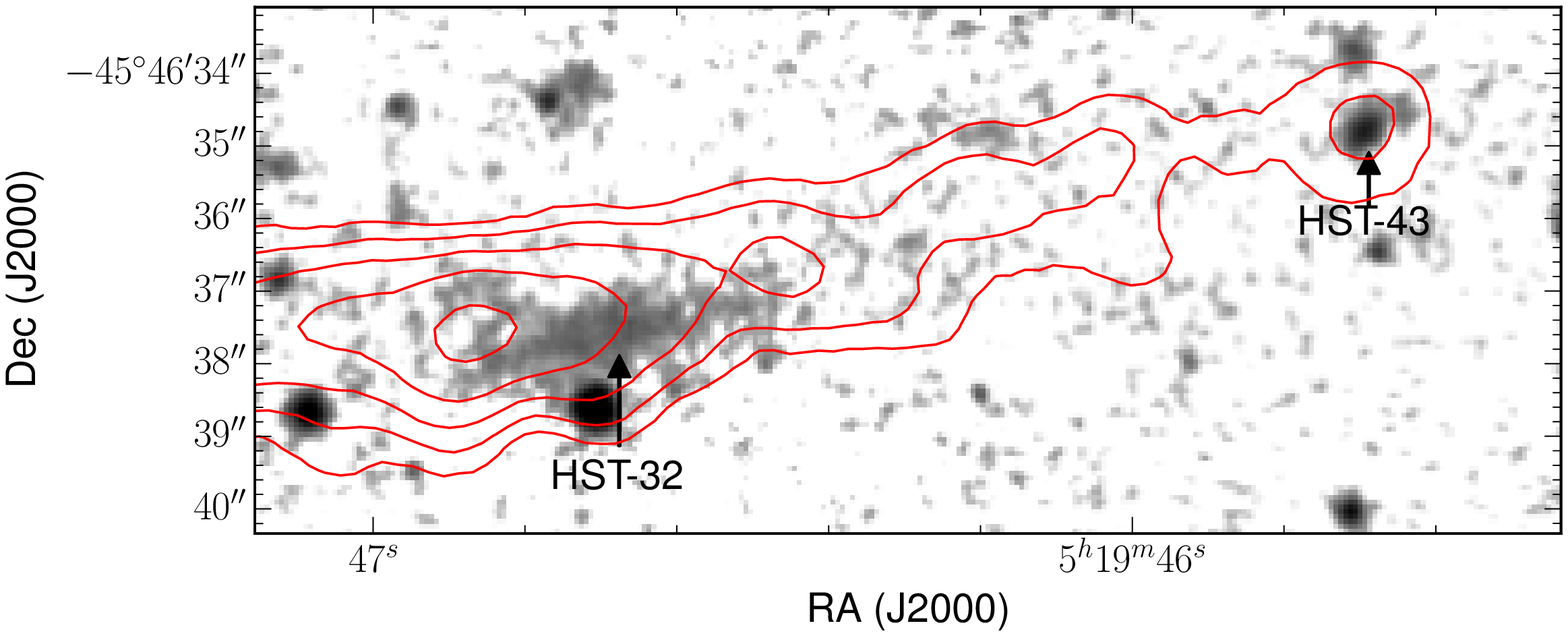}
\includegraphics[width=\columnwidth]{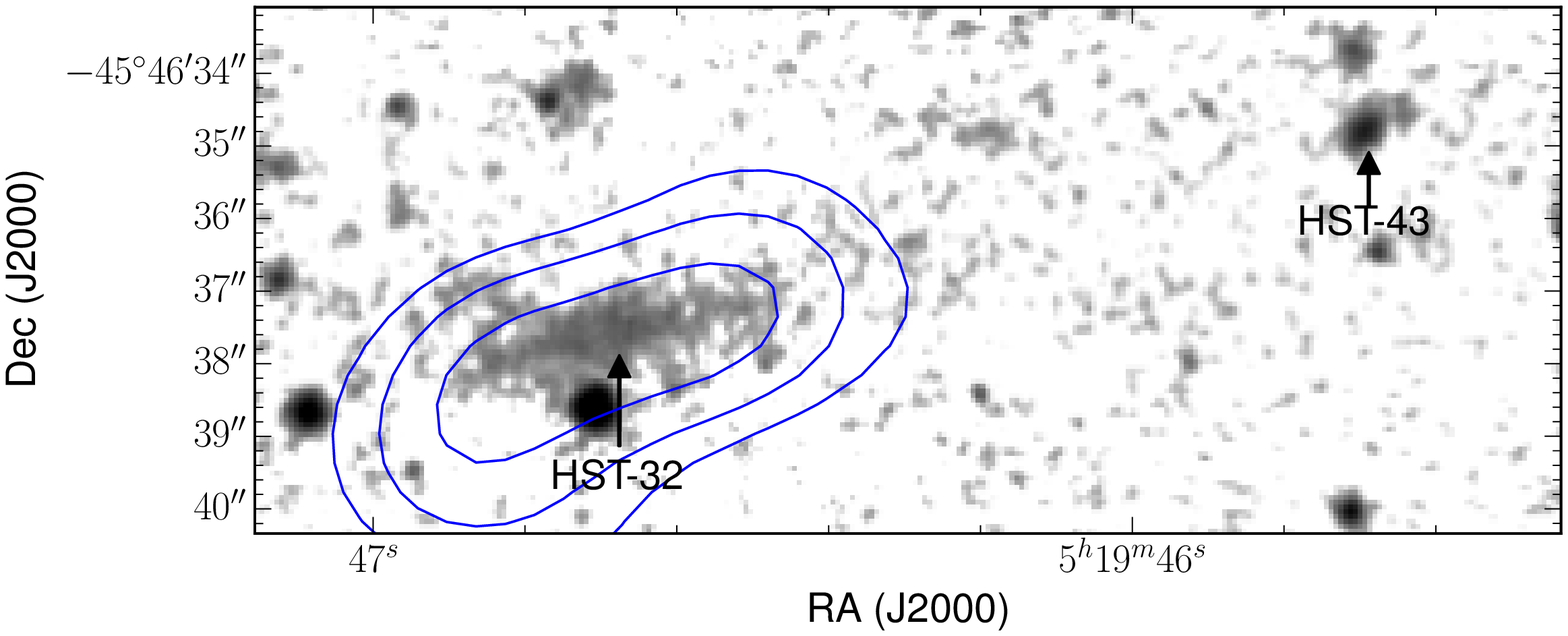}

\includegraphics[width=\columnwidth]{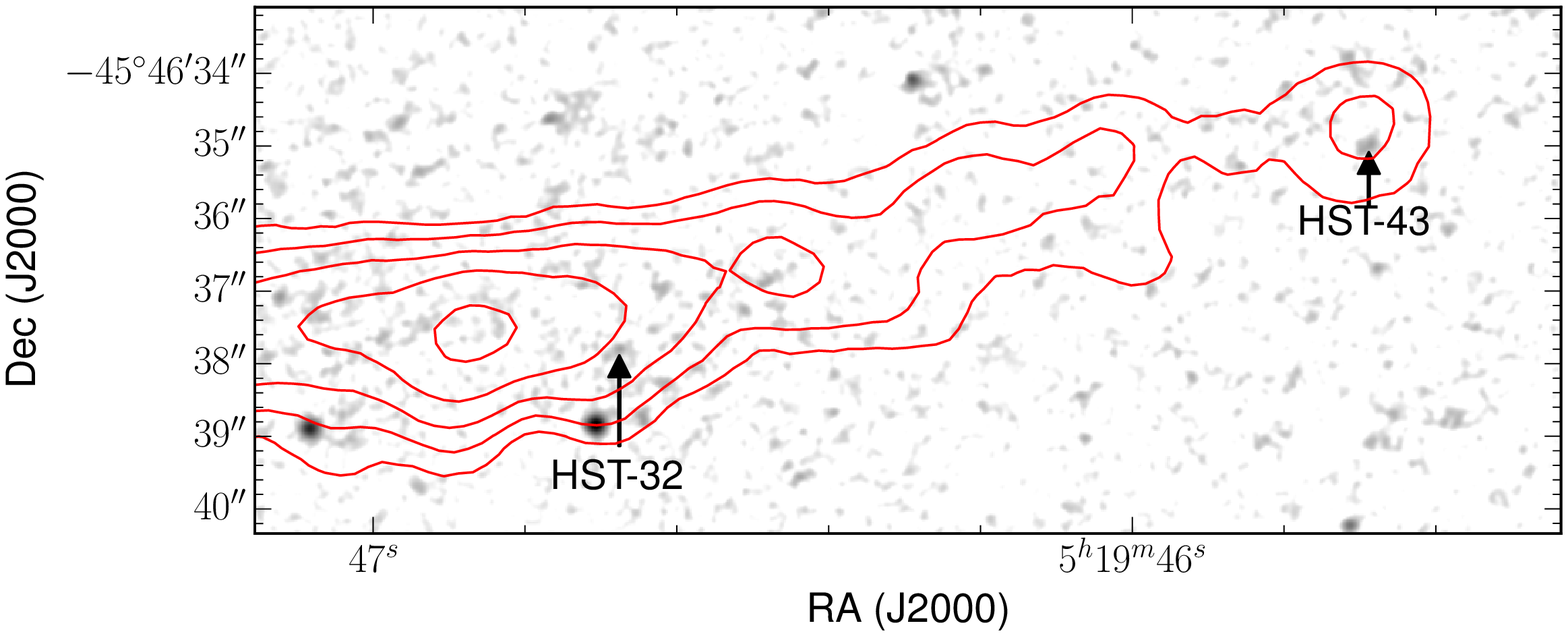}
\includegraphics[width=\columnwidth]{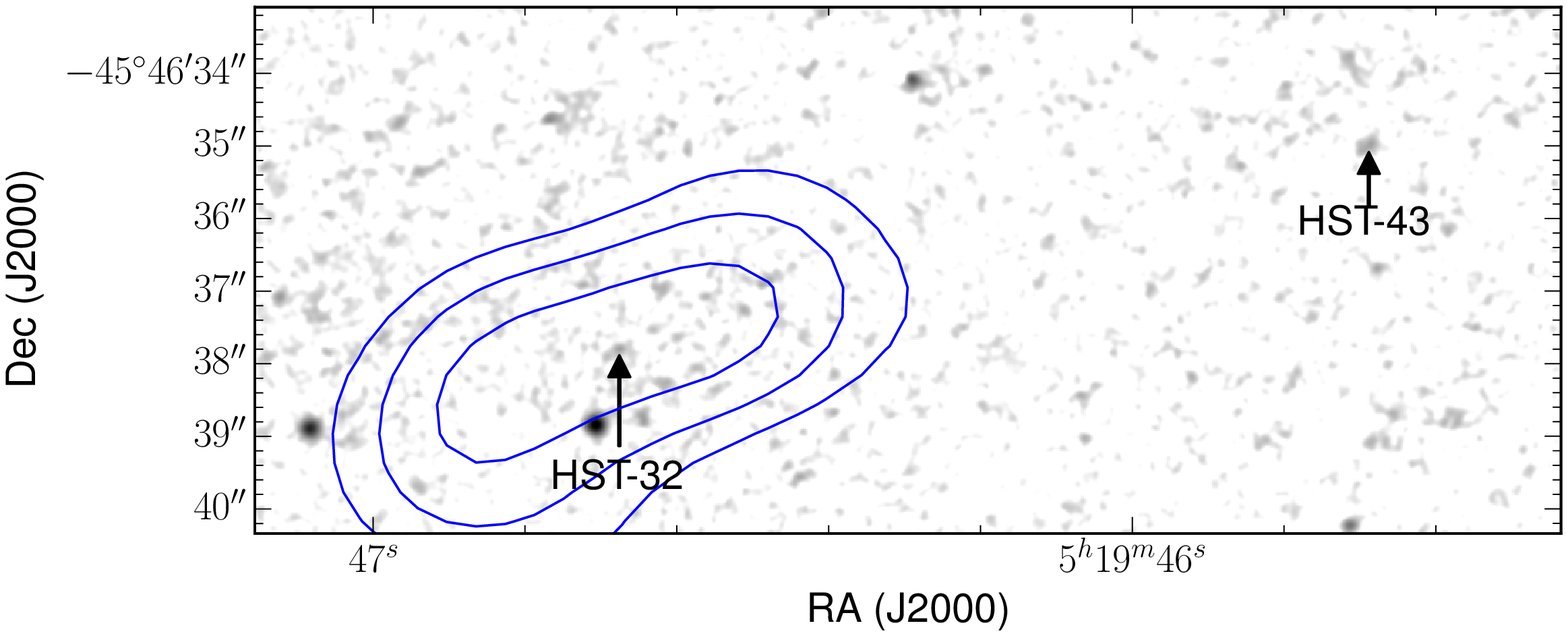}

\includegraphics[width=\columnwidth]{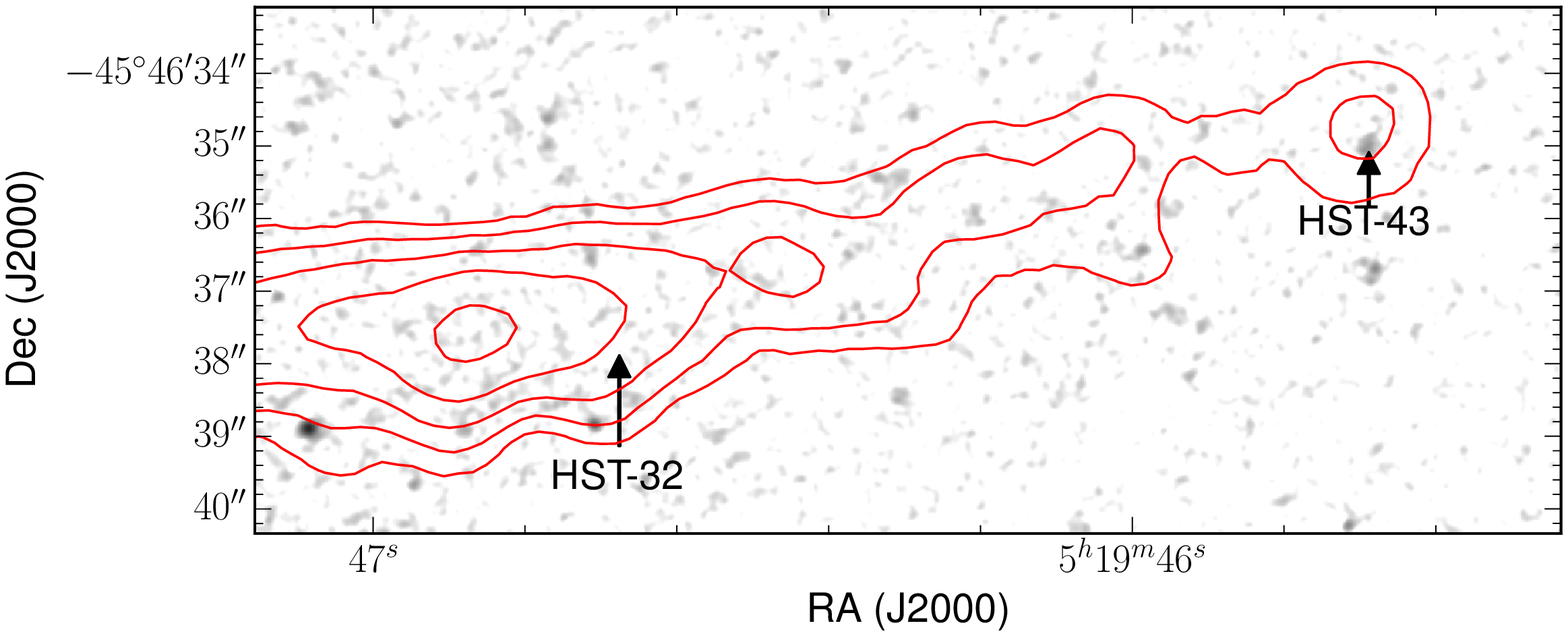}
\includegraphics[width=\columnwidth]{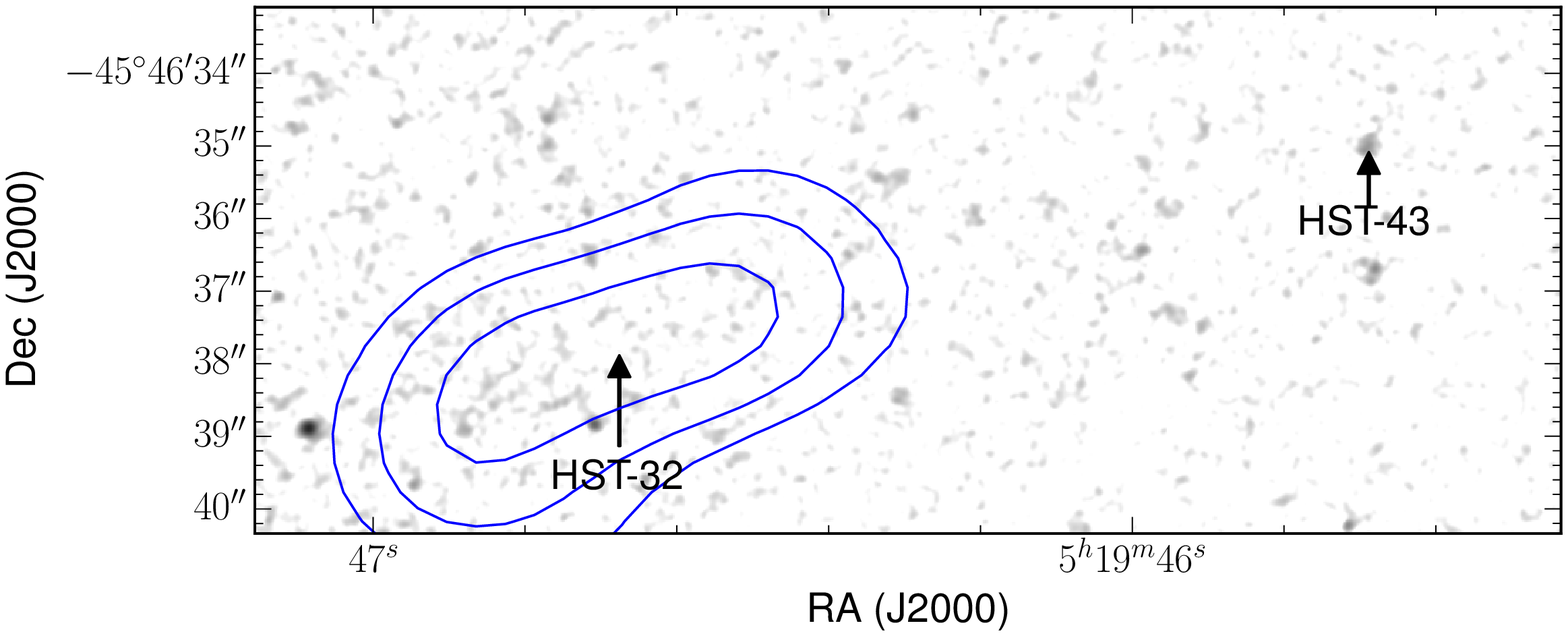} 
  \caption{
  Same as Figure~\ref{fig:jet}, but now cropped to feature knots HST-32 and HST-43.
  HST-32-F160W is shown in more detail in Figure~\ref{fig:gaussian_fits}.}
  \label{fig:inner_knots}
\end{figure*}

\begin{figure*}[tbp]
\centering
\includegraphics[width=\columnwidth]{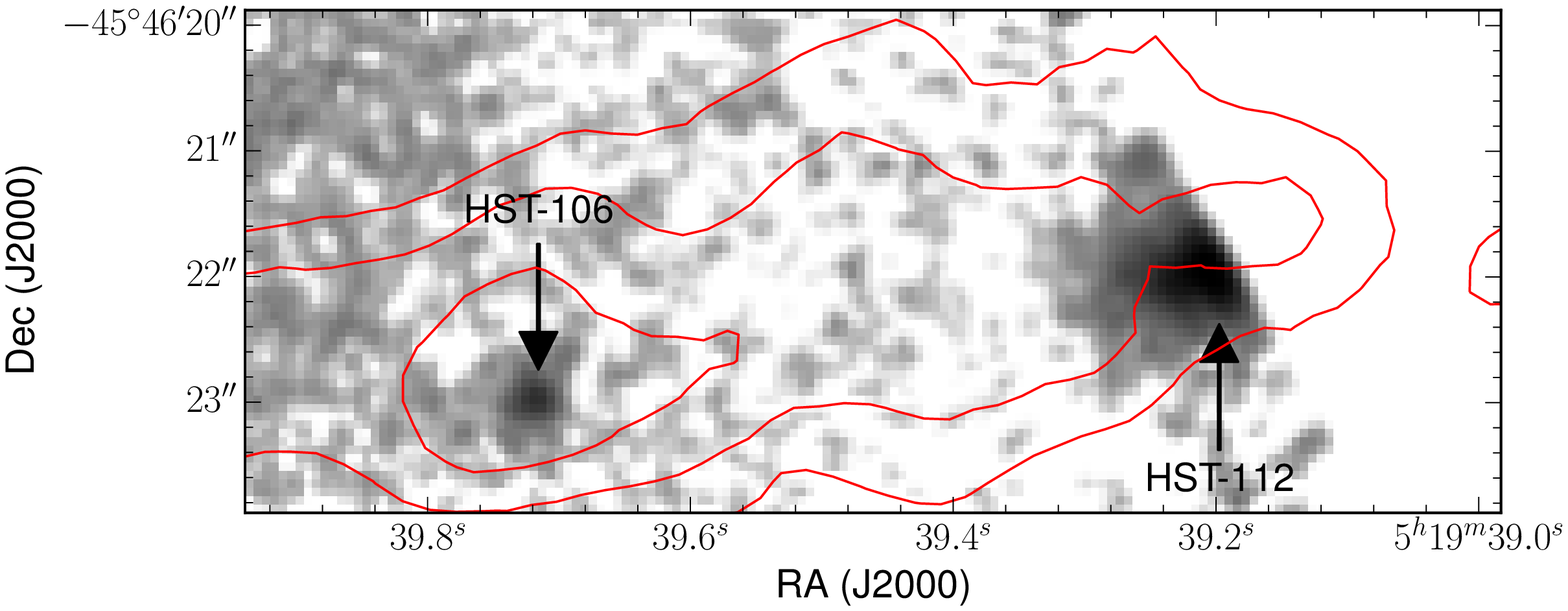}
\includegraphics[width=\columnwidth]{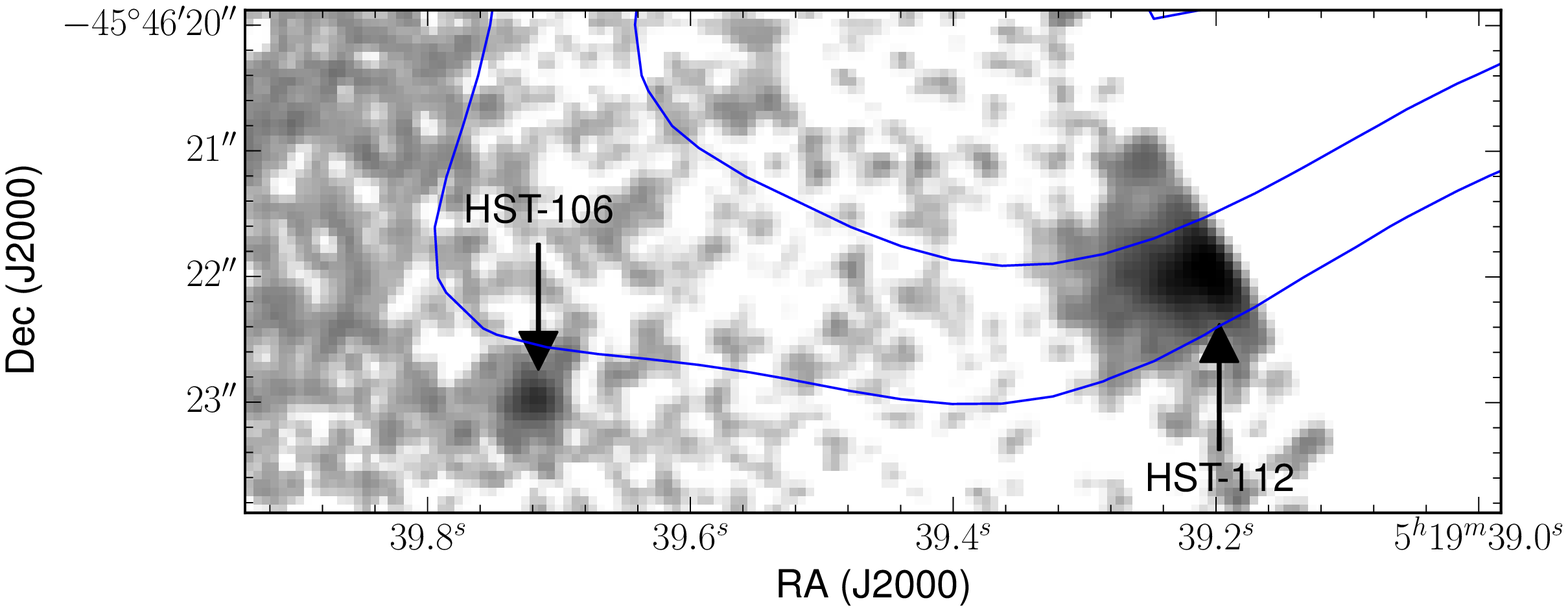}

\includegraphics[width=\columnwidth]{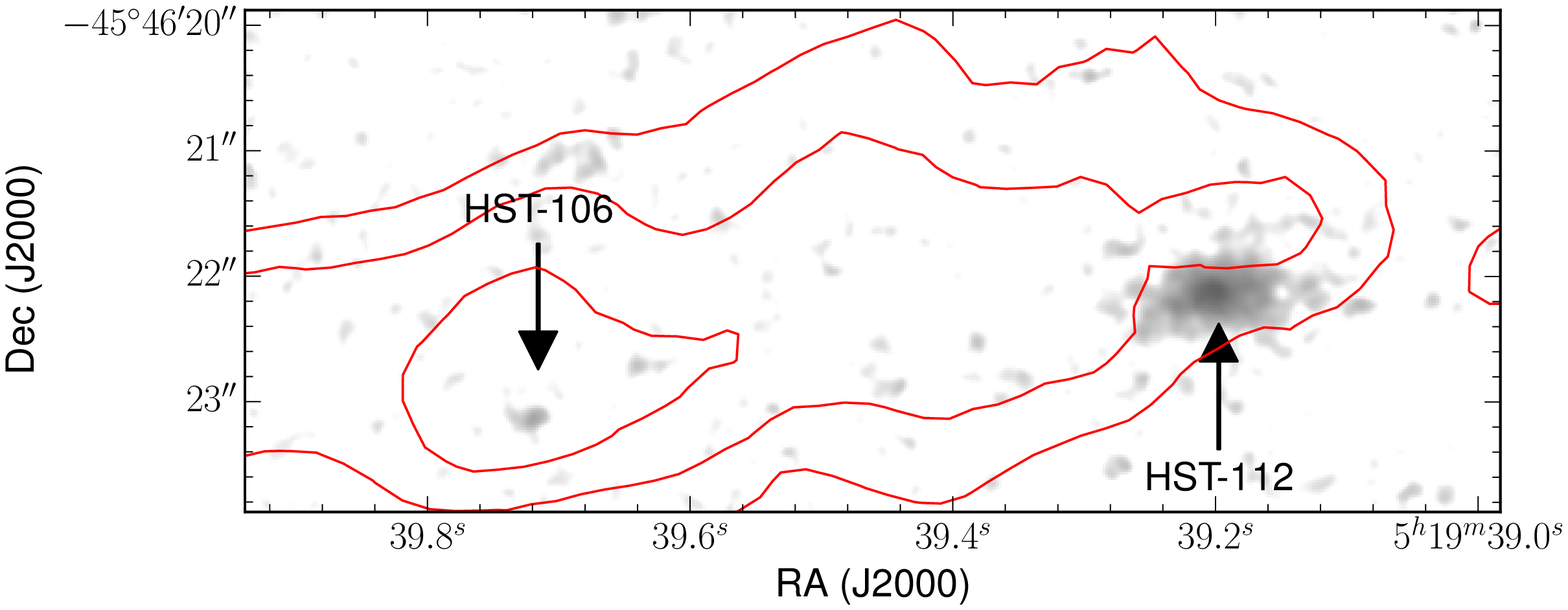}
\includegraphics[width=\columnwidth]{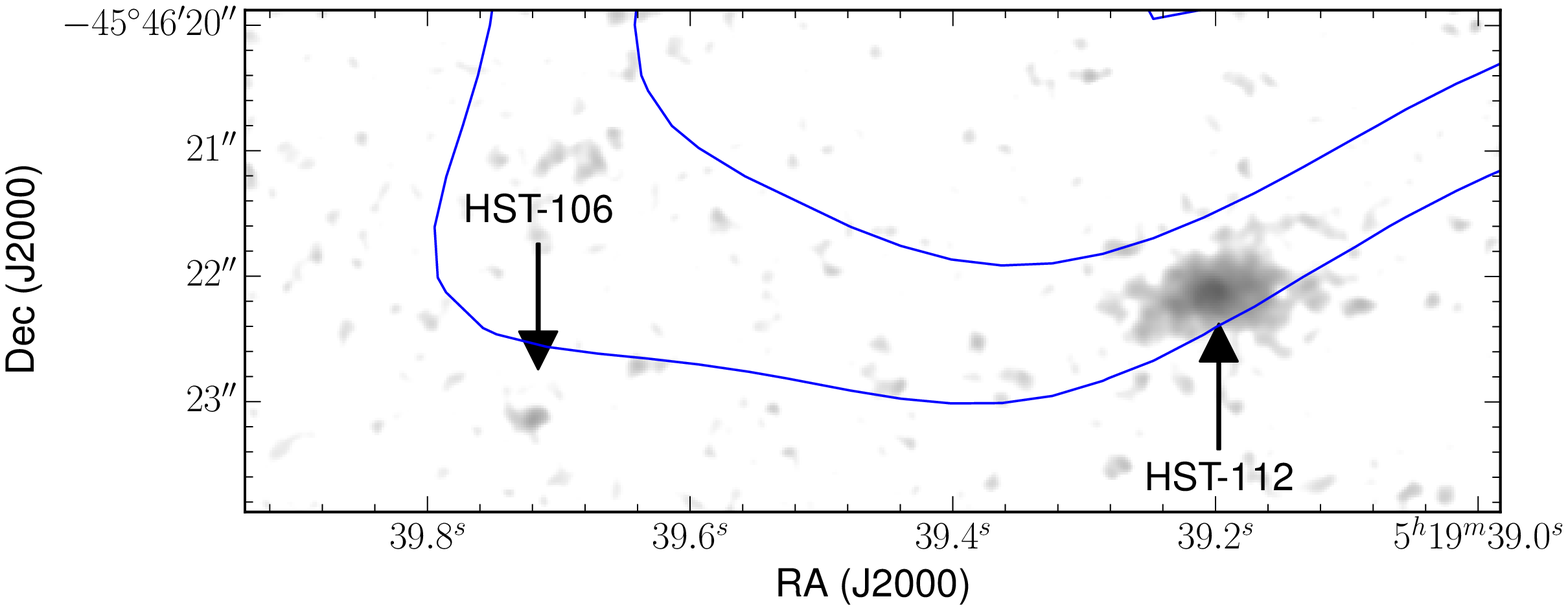}

\includegraphics[width=\columnwidth]{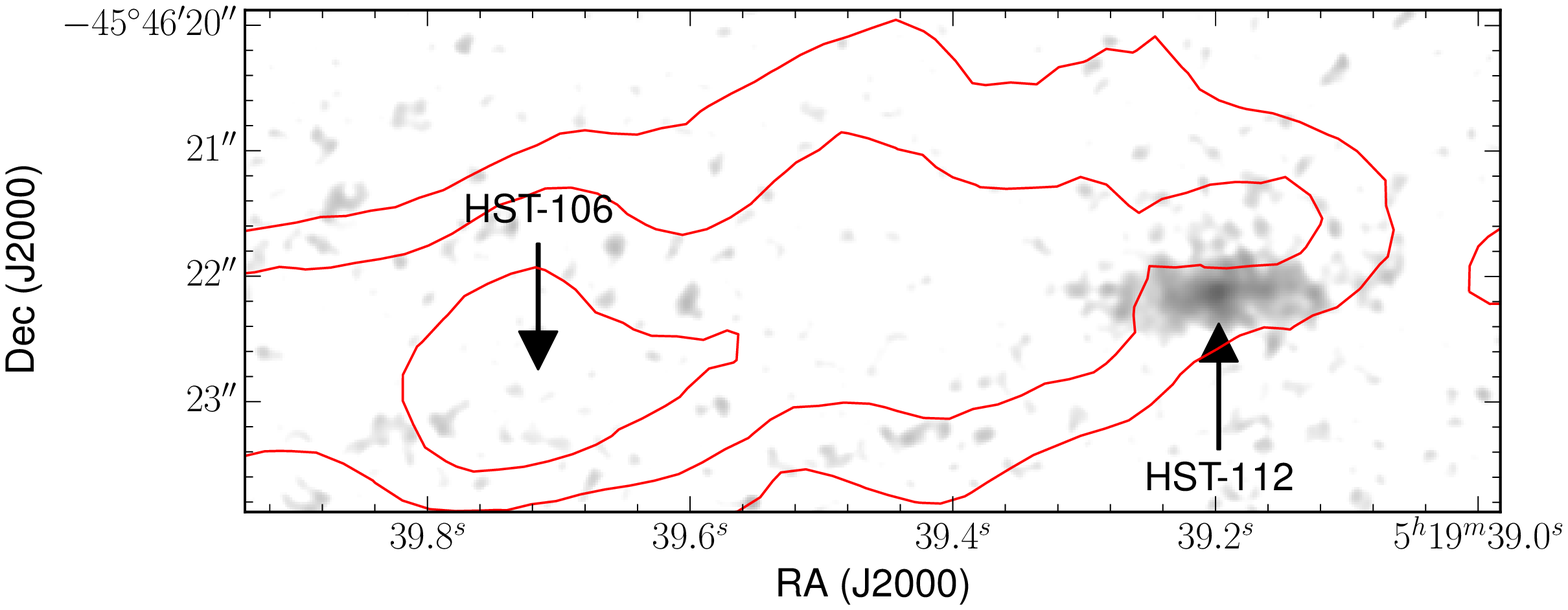}
\includegraphics[width=\columnwidth]{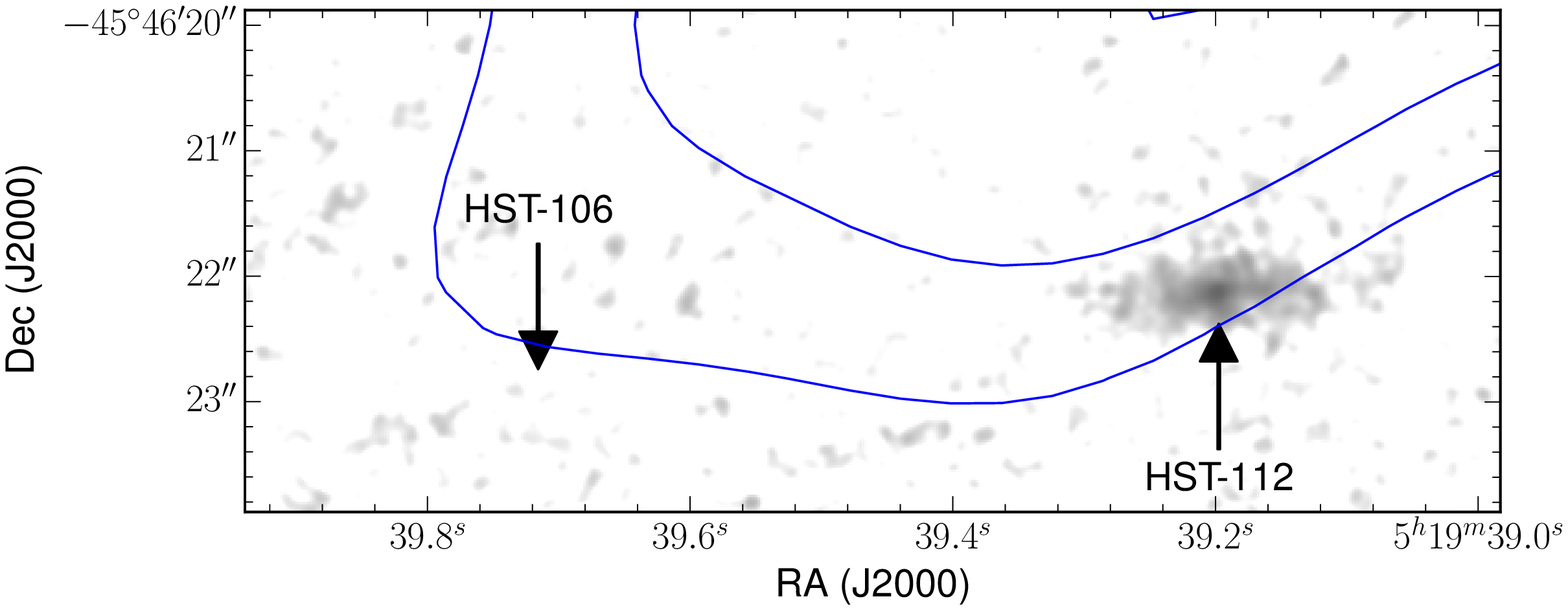} 

\caption{Same as Figure~\ref{fig:inner_knots}, except now for knots HST-106 and HST-112.
  HST-112 lies on the edge of the IR detector and is cut off in the F160W band (top).}
  \label{fig:outer_knots}
\end{figure*}

Four knots were identified, located 32\arcsec, 43\arcsec, 106\arcsec\ and 112\arcsec\ from the core.
We label these knots HST-32 through HST-112 based on their location,
with filter sub-labels appended as necessary (e.g. HST-32-F160W).
These knots are marked on images
Figures~\ref{fig:f160w:galaxy}-\ref{fig:jet}.
Figures~\ref{fig:inner_knots}-\ref{fig:outer_knots} show these knots 
in greater detail, with X-ray and radio contours overlaid.

Overall, the IR image (F160W) is more sensitive 
than the UVIS images (F814W and F475W)
due to the difference in detector sensitivities and exposure times.
Consequently, many of the knots are most clearly detected in the F160W image.

We considered the possibility that what we observe are actually
unrelated background sources, rather than jet knots.
Figure~\ref{fig:jet} shows a large number of optical background sources, 
which could be confused as jet emission. 
By looking for features coincident with observed X-ray flux
within the general jet geometry, 
we reduce the probability of false positives.
\citet{Marshall10} previously ruled out the likelihood that unrelated X-ray background
sources could provide flux densities similar to the ones we observe associated with
these HST knots.  
While we are unable to conclusively determine whether this observed emission 
is produced by the jet, we conclude that the emission is likely from jet knots,
given the correlation in locations and shapes between the X-ray, radio and optical features.

The true extent of the optical component of the jet is probably larger 
than the 4 knots mentioned in this work.
Given the high density of background sources, we have tried to be conservative
in what we claim is a knot.
Other features are likely to become visible with deeper optical imaging.


\subsection{Measuring Jet Knots} 
\label{sec:measuring}

\begin{figure}[tbp]
\centering
\includegraphics[width=\columnwidth]{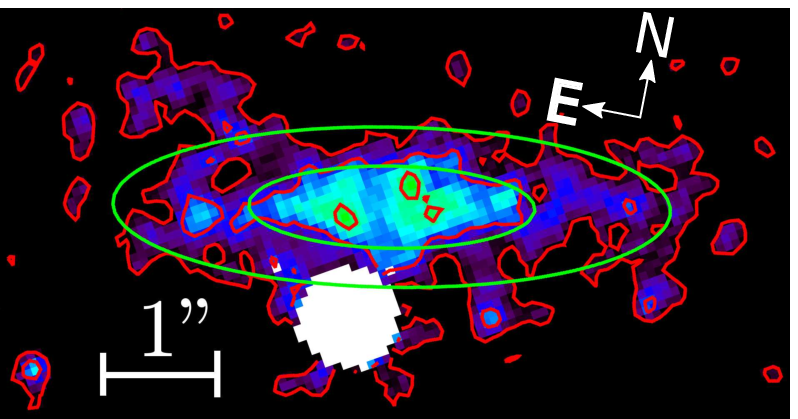}
\caption{HST-32-F160W, along with a 2D Gaussian fit.
	The image has a color scale, inverted in intensity relative to the previous images.
	Red contours are from the observed optical data;
	green contours are from the 2D Gaussian fit to the data.
  Image has been rotated 11\arcdeg\ clockwise, so the jet direction is horizontal,
  going left to right.
	A foreground star has been masked from the analysis (in white).}
\label{fig:gaussian_fits}
\end{figure}

Gaussian 2D fits quantified the location and size of each knot.
Figure~\ref{fig:gaussian_fits} gives a sample visualization of a Gaussian fit applied
to HST-32-F160W, the best detected knot.
Results for all knots and filter images are given in Table~\ref{tab:fittedknots}.
In cases where the fit was only marginally statistically significant, 
we listed the fitted flux density, $S_\nu$, as an upper limit.
Fit visualizations (similar to Figure~\ref{fig:gaussian_fits}) for each knot and band
can be found in the work of \citet[Appendix A]{2014arXiv1405.6704G}, 
but those visualizations are not necessary for this analysis.

These Gaussian fits then determined apertures, 
through which we could extract photometric flux densities.
While we could have used the fitted flux density, 
using rectangular apertures allowed us to be consistent
in how we extracted the radio and X-ray flux densities.  
These apertures were defined to be 2\arcsec\ wide across the jet, 
and approximately 2 Gaussian fit FWHM long in the direction of the jet. 
(Background was subtracted using an adjacent region of identical size.) 
The location and size of each aperture can be found in Table~\ref{tab:apertures}.
In the case of the HST-112-F160W, the desired aperture extends beyond the edge of the image.  
Without knowing how many counts we are missing,  
we simply denote the counts observed as a lower limit.

The Gaussian profiles also allowed us to estimate more accurate upper limits 
for the marginally detected flux densities.
The shape of the F160W fit (clearly detected for all knots) was used 
to create artificial knots of various flux densities, 
which could be injected into the other filter images.  
For each knot and each band, there was a critical flux density 
below which we could not reliably retrieve the artificial knot from the noise.  
We quote that critical value as our upper limits for the aperture flux densities.
These are much more conservative upper limits than those suggested
by the Gaussian fits, which only considered statistical error.
We believe this injection and retrieval process
more accurately reflects the systematic errors which dominate our uncertainties.

This approach of injecting artificial 2D knots would not work for setting limits on the
radio sources, as the radio data were already collapsed into a 1D jet profile before we extracted a flux density.
To assess the significance of a radio detection, 
we estimated a background noise level of approximately 1 mJy beam$^{-1}$ (with a 5 $\mathrm{arcsec}^2$ beamsize)
near the jet (which is significantly higher than the off-source noise level of 40 $\mu$Jy beam$^{-1}$).
While HST-32 and HST-112 are clear detections at this noise level 
(the radio contours of Figures~\ref{fig:jet}-\ref{fig:outer_knots} begin at this noise floor), knots HST-43 and HST-106 are not clearly detected in the radio data, so we denote their flux densities as upper limits.
This introduces uncertainties in the spectral models we will construct in Section~\ref{section:modeling},
but these uncertainties do not affect our final conclusions.

We also considered the possibility of foreground dust extinction affecting these flux densities,
and have found that it plays a negligible role.  Using the work of \cite{2011ApJ...737..103S}
we predict no more than .15 magnitudes (15\%) of extinction in any of these 
optical bands (assuming a galactic column density).
This is sufficiently small to neglect while interpreting our data.


\begin{deluxetable*}{lrrrrrrrrl}
\tablecolumns{8}
\tablewidth{0pc}
\tablecaption{Gaussian Fit Results for Jet Knots \label{tab:fittedknots} }
\tablehead{
     \colhead{Knot Label} & \colhead{Bandpass} & \colhead{$S_\nu$}
         & \colhead{$x$\tablenotemark{a}} & \colhead{$y$\tablenotemark{a}}
     	& \colhead{$s_x$\tablenotemark{b}} & \colhead{$s_y$\tablenotemark{b}}
	& \colhead{PA\tablenotemark{c}} \\
     \colhead{} & \colhead{} & \colhead{($\mu$Jy)}
          & \colhead{(\arcsec)}  & \colhead{(\arcsec)}
          & \colhead{(\arcsec)} & \colhead{(\arcsec)}
          & \colhead{(\arcdeg)} }
\startdata
HST-32 & F160W & $ 2.2 \pm 0.9$ & $32.0 \pm .5$& $ 0.1 \pm .1$ & $3.3 \pm .9$ & $0.9 \pm .3 $ & $1.1 \pm .2$ 
\\
    & F814W & $<$0.18 & \nodata & \nodata & \nodata & \nodata & \nodata 
\\
    & F475W & $<$0.01 & \nodata & \nodata & \nodata & \nodata & \nodata  \\
HST-43 & F160W & $ 0.4 \pm 0.2$ & $42.7 \pm .1$ & $0.6 \pm .1$ & $0.5 \pm .2$ & $0.3 \pm .1$  & $-38.1 \pm .9$ 
\\
    & F814W & $<$0.18 & \nodata & \nodata & \nodata & \nodata & \nodata  \\
    & F475W & $<$0.11 & \nodata & \nodata & \nodata & \nodata & \nodata  \\
HST-106 & F160W & $ 0.2 \pm 0.1$ & $106.4 \pm .8$ & $0.4 \pm .1$  & $0.4 \pm .2$ & $0.3 \pm .2$ & $-52 \pm 3$ 
\\
    & F814W & $<$0.10 & \nodata & \nodata & \nodata & \nodata & \nodata  \\
    & F475W & $<$0.01 & \nodata & \nodata & \nodata & \nodata & \nodata  \\
HST-112 & F160W & $ 3.8 \pm 1.9$ & $112.1 \pm .2$ & $0.4 \pm .1$ & $1.2 \pm .4$ & $0.6 \pm .2$ & $-.1 \pm .3 $ 
\\
    & F814W & $2.2 \pm .2$ & $111.9 \pm .1$ & $0.36 \pm .02$ & $0.85 \pm .06$ & $0.43 \pm .03$ & $1.2 \pm .1$ 
\\
    & F475W & $1.2 \pm .1$ & $112.0 \pm .1$ & $0.39 \pm .01$ & $1.19 \pm .07$ & $0.42 \pm .03$ & $6.0 \pm .1$ 
\\
\enddata
\tablenotetext{a}{Position of Gaussian centroid relative to the core, measured along ($x$)
	or transverse to ($y$) a line with position angle of $-79 \arcdeg$ (E of N) defining the jet.}
\tablenotetext{b}{FWHMs of the 2D Gaussian fits, along ($x$) and transverse to ($y$) the jet.}
\tablenotetext{c}{Position angle of the 2D Gaussian major axis relative to the direction to the core, 
defined to be zero when aligned with the direction to the core and increasing when rotated counter-clockwise.}
\end{deluxetable*}

\begin{deluxetable}{lrrlrrrrr}
\tablecolumns{8}
\tablewidth{0pc}
\tablecaption{Knot Aperture Regions  \label{tab:apertures} }
\tablehead{
     \colhead{Knot Label} & \colhead{RA\tablenotemark{a}}       & \colhead{Dec\tablenotemark{a}}      & \colhead{$s_x$ \tablenotemark{b}} & \colhead{$s_y$ \tablenotemark{c}}  \\ 
      \colhead{}          & \colhead{(hh:mm:ss)} & \colhead{(dd:mm:ss)} & \colhead{(\arcsec)}                & \colhead{(\arcsec) } 
      }
\startdata
HST-32 &   5:19:46.686  & -45:46:37.35 & 6   & 2   \\
HST-43 &   5:19:45.711  & -45:46:34.82 & 1.7 & 2    \\
HST-106 &   5:19:39.732  & -45:46:22.80 & 2   & 2     \\
HST-112 &   5:19:39.197  & -45:46:21.79 & 4   & 2      \\ 
\enddata
\tablenotetext{a}{Box center (J2000)}
\tablenotetext{b}{Box length, along the jet, which is defined to be in the direction -79\arcdeg E of N}
\tablenotetext{c}{Box width, transverse to the jet}
\end{deluxetable}

\begin{deluxetable*}{lrrrrrrrrrr}
\tablecolumns{8}
\tablewidth{0pc}
\tablecaption{Aperture Flux Density Results for Jet Knots  \label{tab:knotfluxes} }
\tablehead{
     \colhead{Knot Label} & \colhead{$S_\nu$(Radio)} & \colhead{$S_\nu$(F160W)} & \colhead{$S_\nu$(F814W)} & \colhead{$S_\nu$(F475W)} & \colhead{$S_\nu$(X-ray)\tablenotemark{a}} & \colhead{$\alpha_\mathrm{X-ray}$\tablenotemark{a}\tablenotemark{b}} \\ 
      \colhead{} & \colhead{(mJy)} & \colhead{($\mu$Jy)} & \colhead{($\mu$Jy)} & \colhead{($\mu$Jy)} & \colhead{(nJy)} & \colhead{}}
\startdata
HST-32	&  	  $3.2$  & $1.98\pm .05$             & $<$ 2.0	        & $< 2.0$       & $1.68 \pm .07$ & $0.96 \pm .07$ 
\\
HST-43	&  	  $<0.2$  & $.26\pm .01$		           & $<$ 0.4	        & $< 0.2$       & $0.14 \pm .06$ & \nodata \tablenotemark{c} 
\\
HST-106	&  	  $<0.4$  & $.22\pm .01$		           & $<$ 0.4          & $< 0.3$       & $0.39 \pm .03$ & $1.12 \pm .15$ 
\\
HST-112	&  	  $1.2$  & $>6.98$\tablenotemark{d}  & $2.47\pm .06$	  & $1.56\pm .04$ & $0.27 \pm .03$ & $0.74 \pm .23$ 

\enddata
\tablenotetext{a}{From the data of M. J. Hardcastle et al.\ (2015, in preparation)}
\tablenotetext{b}{Spectral index, $\alpha$, such that $S_\nu \propto \nu^{-\alpha}$.}
\tablenotetext{c}{Insufficient counts to extract a spectral index; $\alpha=2$ assumed for flux density extraction.}
\tablenotetext{d}{Aperture extended beyond the edge of the detector.}
\end{deluxetable*}


\section{Modeling the Knot Spectra} 
\label{section:modeling}

Using the aperture flux densities of each knot, we tested a number of non-thermal emission models.
Three emission mechanisms were plausible: 
synchrotron radiation from high energy electrons
moving in a magnetic field; 
synchrotron self-Compton emission, 
where synchrotron photons are scattered off of high
energy electrons (SSC); 
and inverse Compton scattering of electrons 
that boosts photons from the cosmic microwave background to
very high energies \citep[IC/CMB; ][]{2000ApJ...544L..23T,2001MNRAS.321L...1C}.
For more details on various models for the physical systems underlying these
emission mechanisms, see the review of \citet{2006ARA&A..44..463H}.
In all models we assume low frequency flux is provided by synchrotron emission;
the label we attach to each model refers to the primary mode of X-ray emission,
with radio and optical emission always produced by synchrotron emission.
In all of the models discussed below, the SSC emission was significantly
smaller than the IC/CMB emission, so we deem SSC emission negligible for the observed knots.
The un-boosted IC/CMB emission was also too low to be shown on our plots,
but we did analyze the amount of boosting that would be required for the
IC/CMB emission to match the observed X-ray flux densities. 

In order to construct these models, we first assumed
the radio and optical flux densities to be from synchrotron radiation.
We need the low energy spectral slope, but since we only have
radio observations in one band, we need to use optical data
to constrain the low energy spectral slope.
From the radio-optical spectral slope, $\alpha$ (such that $S_\nu \propto \nu^{-\alpha}$),
we can then determine the energy index, $p = 2 \alpha + 1$, of the emitting electrons
(assumed to have a power law distribution of energies $N_e(E) \propto E^{-p}$).
Using those values for $p$ (typically $p\sim2$),
the observed radio flux density allows us to determine the magnetic field strength that 
would require the minimal energy in the jet, $B_\mathrm{{me}}$, 
using the equations from \cite{Worrall09}. 
The minimum energy system is approximately equivalent to a system in equipartition.
Table~\ref{tab:modelparams} lists the results for $p$ and $B_\mathrm{{me}}$ for each knot.
It should be noted that $B_\mathrm{me}$ is only accurate to within a factor of 2 or so,
given the assumptions that could be made using the approach of \cite{Worrall09}.

The radio-optical spectral indices differ significantly from the measured
X-ray spectral indices -- there must be a spectral break between the 
optical and X-ray frequencies, a \emph{softening} of the spectrum.
The most likely option is that there is a break in the electron energy distribution itself.
(For a continuous, but kinked distribution we will use the label \emph{broken};
for a distribution with a jump discontinuity we will use the label \emph{multiple populations}.)
A break in electron energies is predicted by the continuous injection model: 
if electrons are injected with a single power law of energies, the highest energy electrons
would experience the greatest radiative losses,
resulting in a steady state distribution composed of a broken power law, 
with $p_\mathrm{high} \approx p_\mathrm{low} + 1$ \citep[Section~16.3.1]{2011hea..book.....L}.
This is precisely our situation, if we assume all of the observed emission is coming from 
synchrotron processes. We infer $p_\mathrm{low}\sim2$ 
for the lower-energy electrons responsible for the radio-optical emission, 
and $p_\mathrm{high}\sim3$ for the higher-energy electrons 
responsible for the X-ray emission.

This spectral break can also be used to gain information on the age of a knot.
For an electron with a Lorentz factor $\gamma$, in a magnetic field of strength $B$
we expect the electron to have a characteristic synchrotron cooling time scale of:
\begin{equation}
  t_\mathrm{cool} = 2.4 \cdot 10^7 \left( \frac{B}{\mu\mathrm{G}}\right)^{-2} \left( \frac{\gamma}{10^6} \right)^{-1} \; \mathrm{ yr}
\end{equation}
\citep[adapted from][]{2011hea..book.....L}. 
Evaluated at the break energy, $\gamma_\mathrm{break} m_e c^2$, we get a characteristic
time scale $t_\mathrm{break}$.
For the continuous injection model, this characteristic time scale is expected to be 
equal to how long an injection event has been continuously occurring for a particular knot.
For a model with multiple electron populations, this time scale corresponds to the maximum time that 
could have elapsed since the low energy population was last accelerated.
(The high energy population must have been accelerated even more recently, given the shorter time scales
at higher $\gamma$.)

\begin{figure}[tbp]
\includegraphics[width=\columnwidth]{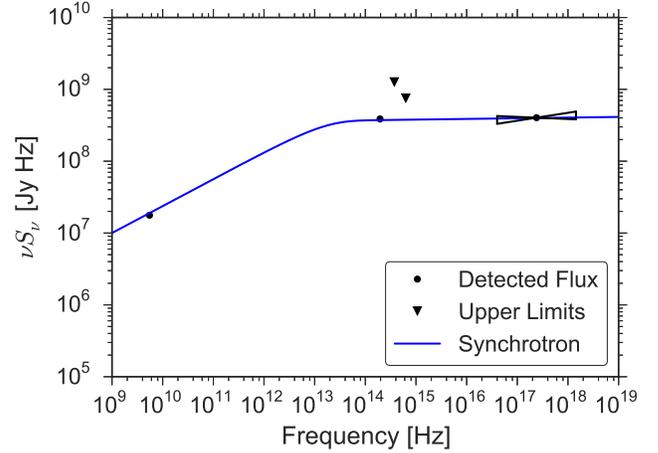}
  \caption{SED model for HST-32. 
  Flux densities are listed in Table~\ref{tab:knotfluxes},
  and model components are listed in Table~\ref{tab:modelparams}.  }  
\label{fig:SED:HST1}
\end{figure}

\begin{figure}[tbp]
\includegraphics[width=\columnwidth]{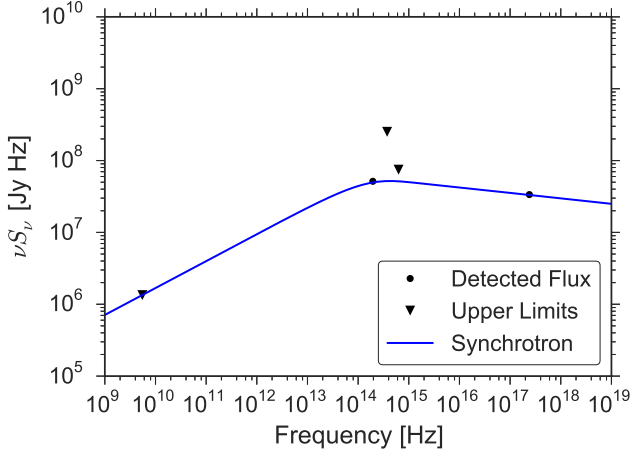}
  \caption{Same as Figure~\ref{fig:SED:HST1},
  but for HST-43.  
\label{fig:SED:HST2} }
\end{figure}

\begin{figure}[tbp]
\includegraphics[width=\columnwidth]{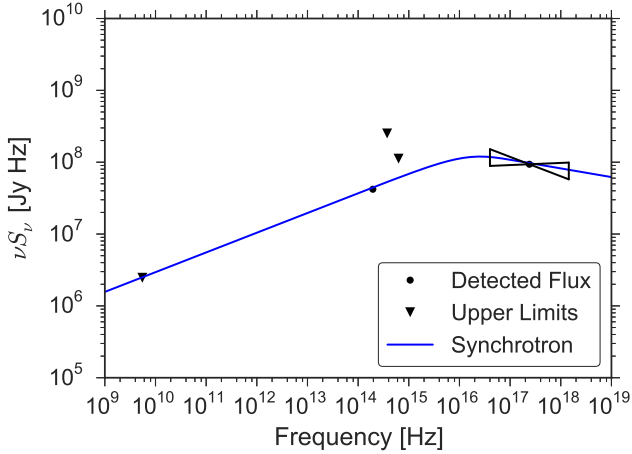}
  \caption{Same as Figure~\ref{fig:SED:HST1},
  but for HST-106.
\label{fig:SED:HST3} }
\end{figure}

\begin{figure}[tbp]
\includegraphics[width=\columnwidth]{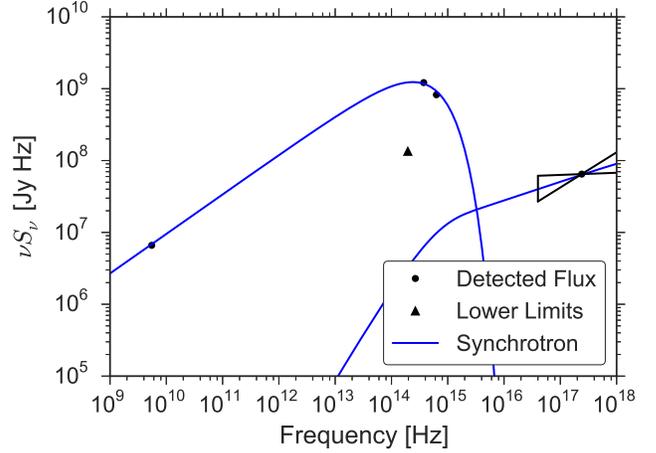}
  \caption{Same as Figure~\ref{fig:SED:HST1},
  but for HST-112.
  A single broken power law was insufficient; a second power law (with different slope and normalization) was needed.
\label{fig:SED:HST4} }
\end{figure}

For each knot we constructed an SED consisting of synchrotron emission
from a broken power law of electron energies.
The low energy indices were set by the radio and optical flux densities,
while the high energy spectral indices were set by the observed X-ray spectral indices.
If possible, a continuous but kinked electron energy distribution was used.
If that was not possible (i.e. HST-112), we used 2 synchrotron spectra with differing normalization
(a \emph{multiple population} model, introducing a discontinuity in the spectrum).
The results for each knot can be seen in
Table~\ref{tab:modelparams} and Figures~\ref{fig:SED:HST1}-\ref{fig:SED:HST4}.

These results can be contrasted with models which are used to explain the
X-ray emission through IC/CMB emission.
Either through bulk Doppler boosting or lowered magnetic field strengths,
IC/CMB X-ray emission can be boosted relative to the 
lower-energy synchrotron emission.
In the case of bulk Doppler boosting, 
we can use the X-ray flux density predicted by an un-boosted model to infer the required Doppler factor: 
$\delta_\mathrm{req}^{1+\alpha} = S_{\nu, \mathrm{observed}} / S_{\nu, \mathrm{predicted}}$,
where $\delta = \left( \Gamma \left[1 - \beta \cos\theta \right] \right)^{-1}$ 
for a bulk Lorenz factor $\Gamma$ and an angle $\theta$ between the jet and the line of sight
\citep{Worrall2006}.
This approach for X-ray emission only works for values of $\delta < 10^3$; 
beyond that limit, assumptions in the scaling break down.
All of our models are close to, or beyond this limit, 
indicating that unphysical levels of Doppler boosting would be required.
A similar approach can be done for lowered magnetic field strengths,
following the scaling relation:
$(B_\mathrm{req} / B_\mathrm{me})^{-(1+\alpha)} = S_{\nu, \mathrm{observed}} / S_{\nu, \mathrm{predicted}}$,
with the total energy (in the case of lowered magnetic field strengths) following the relation:
$E_\mathrm{req} / E_\mathrm{me} \approx (B_\mathrm{req} / B_\mathrm{me})^{-(\alpha+1)}$.
These scaling relations would require magnetic field strengths
around $10^3$ below equipartition, with an energy
roughly $10^{4.5}$ times greater than the minimal energy. We consider such a configuration unphysical.

Even if the amount of IC/CMB X-ray emission could be significantly boosted,
we would expect that the spectral index would be identical to that of the
radio-optical band, whereas our data rule out this possibility.
One might argue that in addition to unphysical amounts of boosting required,
there might also be a break in the electron energy distribution
between the IC/CMB emitting electrons (lower energy)
and the synchrotron emitting electrons (higher energy),
but that would imply a \emph{hardening} of the energy distribution
($p_\mathrm{high} \approx p_\mathrm{low} - 1$).
Such a configuration would be unusual---the opposite of the softening predicted by
the continuous particle injection model---and is difficult to explain.
(The work of \citet{2004ApJ...608..698S} would have tentatively labeled 
Pictor A's jet a synchrotron emitter on the grounds of our spectral data as well; see their Figure 4.)
Between the unphysical levels of boosting
and the unexplained hardening of the electron energy spectrum
required for a boosted IC/CMB model, 
we rule out boosted IC/CMB as the primary emission mechanism
for the X-ray component of Pictor A's jet.


\begin{deluxetable*}{lrrrrrrrrr}
\tablecolumns{4}
\tablewidth{0pc}
\tablecaption{Synchrotron Model Results\tablenotemark{a} \label{tab:modelparams} }
\tablehead{
     \colhead{Knot Label} & \colhead{$p_\mathrm{low}$} & \colhead{$\alpha_\mathrm{low}$} 
     & \colhead{$\gamma_\mathrm{break}$} &  \colhead{$p_\mathrm{high}$} & \colhead{$\alpha_\mathrm{high}$}
     & \colhead{$B_\mathrm{me}$} & \colhead{$\delta_\mathrm{req}$} & \colhead{$t_\mathrm{break}$}
     \\
     \colhead{} & \colhead{} & \colhead{}  & \colhead{($10^6$)} & \colhead{} & \colhead{} & \colhead{($\mu$G)}  & \colhead{} & \colhead{ (yr)} }
\startdata
HST-32 & 2.25     & 0.62  & 39 &  2.98 & 0.99  & 33 & $4 \cdot 10^2$ & 576
\\ 
HST-43 & $<2.25$     & $<0.62$  & 26 &  3.15 & 1.08  & 49 & $<3 \cdot 10^3$ & 392
\\ 
HST-106 & $<2.45$     & $<0.72$  & 21 &  3.24 & 1.12  & 88 & $<2 \cdot 10^3$ & 150
\\
HST-112 & 1.90     & 0.45  & 21 &  2.48 & 0.74  & 29 & $3 \cdot 10^3$ & 1388
\enddata
\tablenotetext{a}{See \S~\ref{section:modeling} for definitions of these parameters of the synchrotron model
fits to each knot.}
\end{deluxetable*}


\section{Tidal Tail}
\label{section:tidal_tail}

A tidal tail, clearly visible in Fig.~\ref{fig:f160w:resid}, was an unanticipated find.
This tail starts coincident with a radio source 18\arcsec\ (12 kpc projected distance) north of Pictor A,
sweeping to the west along a path 90\arcsec\ (60kpc) long, 
and ending 90\arcsec\ (60kpc) from the center of Pictor A.
(Many galaxy subtraction variants were attempted,
but none showed the tail closer than the radio source 18\arcsec\ from the center of Pictor A.)
At its closest to Pictor A, the tidal tail is approximately
3\arcsec\ (2 kpc) wide, becoming at least 7\arcsec\ (5 kpc) wide
at its farthest from Pictor A.  The tail was only detected in
our deepest image (the F160W data),
but we expect deeper observations would reveal it in other optical bands as well.

The classification scheme of \cite{2010ApJ...709.1067B} suggests that
this long tidal tail is well into the first passage of a merger.
Dynamical estimates predict such a tail would take a few hundred million
years to form, and would last another few hundred million years.
This is certainly long enough for the formation of a jet at least 160 kpc long,
suggesting a possible connection between Pictor A's large galactic-scale jet
and a recent merger with the coincident radio source.

\section{Discussion \& Conclusions } 
\label{section:conclusion}

Our data contain the first optical detection of Pictor A's jet, but we only
clearly detect at most a few knots rather than the entire length of the jet.
Although these data are sparse, they allow us to construct physical models
of the jet emission at multiple points along its extent.

The first major result was these optical data give
radio-optical spectral slopes, $\alpha$, which differ significantly
from the X-ray spectral indices measured by M. J. Hardcastle et al.\ (2015, in preparation).
This discrepancy rules out a single, unbroken synchrotron spectrum.
It also makes the boosted IC/CMB hypothesis very unlikely.

Our optical data show that two components of synchrotron emission 
are much more consistent with observations.
The break between these components would results in a softening
of the electron energy distribution, around $\gamma\sim 10^6$.  
Such a softening would be a natural
consequence of a continuous injection model, where electrons are constantly
being accelerated to an initial power law of energies, but radiative losses
result in a steady state broken power law of energies.
The injection time scales implied by these spectral breaks
are quite short, in the hundreds of years, while the light-crossing
time scale is at least 1000 yr for each knot.
We caution that deeper optical observations could rule out this
single population synchrotron model with a cooling break,
at which point this spectral break time scale would no longer give the
duration of an ongoing injection event. 
If the single population synchrotron model were ruled out,
a multiple population model might be necessary, which was the case for knot HST-112.

Ruling out a boosted IC/CMB model is be a key element to understanding
the emission of Pictor A's jet.  
The work of \cite{Hardcastle05} provided independent evidence 
against an IC/CMB model for Pictor A's jet emission, which has
recently been confirmed by M. J. Hardcastle et al.\ (2015, in preparation).
Our work, providing optical data, also reaches the conclusion
that a boosted IC/CMB model for Pictor A's jet emission
does not match observed data.

Still unconfirmed is the possibility of unresolved
regions hypothesized by \citet{Marshall10} 
based on X-ray variability data.
Unfortunately, our images were insufficient for identifying unresolved regions
definitively associated with knot emission. 
All currently detected optical knots are resolved on scales similar to the width
of the X-ray jet emission.
Deeper observations might yet uncover the unresolved regions hypothesized by \citet{Marshall10}.

Deeper observations would also help constrain our SEDs.
While most of the UVIS knots are non-detections
(HST-32 through HST-106, in the F814W and F475W images),
our detection of UVIS flux from HST-112 allows us to put
stronger constraints on emission models. 
Deeper observations would help us understand if the other SEDs
are simply from a single population broken electron power law
(the simplest explanation) or from two distinct electron
populations (such as HST-112).  
If the underlying electron distributions are broken, 
but continuous (as expected for a single population continuous injection model),
determining the break energy would help constrain the 
time scales of particle acceleration events along the length of the jet.
If instead the electrons are from multiple populations
(a broken and discontinuous distribution),
the high energy population might also help us better understand 
how regions of Pictor A's jet could flare on short time scales,
while emission from the majority of the jet remained stable.
Better data, and stronger limits, would help us answer these questions.

Finally, we have noted that our data contain the first evidence
for a tidal tail near Pictor A, 
but it has only been detected in the F160W band.  
Deeper observations in visible bands 
would allow the study of this tidal tail
and its role in the dynamics of Pictor A.

\acknowledgements

We thank the referee for useful feedback which helped improve 
the quality of this work.
Support for this work was provided by the National Aeronautics and
Space Administration through the Smithsonian Astrophysical Observatory
contract SV3-73016 to MIT for Support of the Chandra X-Ray Center,
which is operated by the Smithsonian Astrophysical Observatory for and
on behalf of the National Aeronautics Space Administration under contract
NAS8-03060.

The Centre for All-sky Astrophysics (CAASTRO) is an Australian Research Council Centre of Excellence, funded by grant CE110001020.

\bibliographystyle{apj}                       
\bibliography{opticaljet}

\end{document}